\documentclass{ieeeaccess}
\usepackage{cite}
\usepackage{amsmath,amssymb,amsfonts}
\usepackage{graphicx}
\usepackage{textcomp}
\usepackage{hyperref}					
\hypersetup{colorlinks,
	linkcolor=black,%
	citecolor=black}

\usepackage{subfigure}
\usepackage{setspace}
\usepackage{url}
\usepackage{color}

\usepackage{algorithm}
\usepackage{algorithmicx}
\usepackage{algpseudocode}

\usepackage{tabularx}

\def\BibTeX{{\rm B\kern-.05em{\sc i\kern-.025em b}\kern-.08em
    T\kern-.1667em\lower.7ex\hbox{E}\kern-.125emX}}
\begin{document}
\history{Date of publication xxxx 00, 0000, date of current version xxxx 00, 0000.}
\doi{10.1109/ACCESS.2023.0322000}

\title{An Intelligent SDWN Routing Algorithm Based on Network Situational Awareness and Deep Reinforcement Learning.}
\author{\uppercase{JINQIANG LI}\authorrefmark{1}, \uppercase{MIAO YE}\authorrefmark{1,2}, \uppercase{LINQIANG HUANG}\authorrefmark{3},\uppercase{XIAOFANG DENG}\authorrefmark{1,2}, \uppercase{HONGBING QIU}\authorrefmark{1} and \uppercase{YONG WANG}\authorrefmark{3,4}.}

\address[1]{School of Information and Communication Technology, Guilin University of Electronic Technology,Guilin 541000,China}
\address[2]{Guangxi Key Laboratory of Wireless Broadband Communication and Signal Processing, Guilin 541004, China}
\address[3]{School of Computer Science and Information Security, Guilin University of Electronic Technology, Guilin 541004, China}
\address[4]{Engineering Technology Research Center of Cloud Security and Cloud Service, Guilin University of Electronic Technology, Guilin 541004, China}

\tfootnote{This work was supported in part by the National Natural Science Foundation of China (Nos.62161006, 62172095, 61861013), the subsidization of the Innovation Project of Guangxi Graduate Education (Nos. YCSW2022270, YCSW2023310, YCBZ2023134), and Guangxi Key Laboratory of Wireless Wideband Communication and Signal Processing (No. GXKL06220110).}

\markboth
{Author \headeretal: Preparation of Papers for IEEE TRANSACTIONS and JOURNALS}
{Author \headeretal: Preparation of Papers for IEEE TRANSACTIONS and JOURNALS}

\corresp{Corresponding author: XIAOFANG DENG(e-mail: xfdeng@guet.edu.cn).}

\begin{abstract}
Due to the highly dynamic changes in wireless network topologies, efficiently obtaining network status information and flexibly forwarding data to improve communication quality of service are important challenges. This article introduces an intelligent routing algorithm (DRL-PPONSA) based on proximal policy optimization deep reinforcement learning with network situational awareness under a software-defined wireless networking architecture. First, a specific data plane is designed for network topology construction and data forwarding. The control plane collects network traffic information, sends flow tables, and uses a GCN-GRU prediction mechanism to perceive future traffic change trends to achieve network situational awareness. Second, a DRL-based data forwarding mechanism is designed in the knowledge plane. The predicted network traffic matrix and topology information matrix are treated as the environment for DRL agents, while next-hop adjacent nodes are treated as executable actions. Accordingly, action selection strategies are designed for different network conditions to achieve more intelligent, flexible, and efficient routing control. The reward function is designed using network link information and various reward and penalty mechanisms. Additionally, importance sampling and gradient clipping techniques are employed during gradient updating to enhance convergence speed and stability. Experimental results show that DRL-PPONSA outperforms traditional routing methods in network throughput, delay, packet loss rate, and wireless node distance. Compared to value-function-based Dueling DQN routing, the convergence speed is significantly improved, and the convergence effect is more stable. Simultaneously, its consumption of hardware storage space is reduced, and efficient routing decisions can be made in real-time using the current network state information. The source code can be accessed at https://github.com/GuetYe/DRL-PPONSA.
\end{abstract}


\titlepgskip=-21pt

\maketitle

\section{Introduction}
\label{sec:introduction}
\PARstart{W}{ith} the rapid development of wireless local area network (WLAN) technology and the exponential growth in the number of mobile terminals in operation, wireless access points (APs) are being provided in more locations to allow users to access the internet. This has resulted in explosive growth in wireless network traffic. In addition, the bursty traffic and rapid mobility of mobile wireless users can cause significant delay and packet loss issues in traditional wireless networks. This situation presents significant challenges for traditional wireless network architectures in terms of the physical infrastructure, protocol framework, and transmission performance, severely affecting the communication quality of service (QoS) for wireless users \cite{b1}. The efficient routing and transmission of data in wireless networks rely on the timely and flexible acquisition of network state information and the design of efficient routing algorithms. Therefore, researching ways to optimize the wireless network architecture and routing protocol design to improve transmission efficiency has important theoretical significance and practical value for maximizing the utilization of network resources in current wireless transmission scenarios.

In traditional wireless network frameworks, data forwarding and control management are tightly coupled, which limits network scalability. Consequently, network devices need to support multiple integrated network functions, leading to a cumbersome network architecture that hinders network management and maintenance \cite{b2}. Moreover, traditional distributed network architectures have deployment limitations. In the deployment of new network services in large-scale dynamic networks, complex and heterogeneous network structures can easily arise. To address these issues, software-defined wireless networking (SDWN) is considered a reliable solution that aims to improve the programmability and flexibility of wireless network architectures to facilitate more efficient network management and performance optimization. SDWN is an emerging wireless network framework that is an extension of software-defined networking (SDN). In this framework, the vertical structure of traditional wireless networks is broken up by separating the complex control functions of traditional wireless network devices and logically concentrating them on controllers to decouple data forwarding and control management. In this way, direct programmability and centralized control of network logic are realized along with the abstraction of the underlying wireless infrastructure, enabling network management through software. An SDWN controller schedules network resources from a global perspective and at a fine-grained level, using the southbound interface of the OpenFlow protocol and the data plane for programming control. Thus, the controller achieves the functions of issuing flow tables and collecting network status information. Meanwhile, the northbound interface of the controller provides an open interface with the application plane \cite{b3}. Based on the advantages outlined above, data forwarding and control management can be decoupled in SDWN, providing the possibility for routing algorithms to reschedule and efficiently utilize network resources.

Efficient routing protocol design is crucial in wireless networks, and the optimization of routing algorithms is a key area of research in traffic engineering \cite{b4}. A routing algorithm aims to find the most efficient routing strategy from the source node to the destination node in the network topology. The algorithm's performance determines the QoS in the wireless network. Consequently, designing a routing algorithm that is both efficient and stable is crucial for enhancing overall performance and resource utilization in an SDWN environment.

In recent years, researchers have enhanced network performance by optimizing routing algorithms. Traditional routing algorithms, such as distance vector routing protocols (DVRPs) \cite{b5} and open shortest path first (OSPF) algorithms \cite{b6}, are primarily based on the shortest path approach. However, these algorithms often use only limited link information for optimization and cannot fully utilize the global network information to enhance network performance \cite{b3}. Moreover, traditional routing algorithms have shortcomings such as a slow convergence speed, a long response time, and difficulty adapting to dynamic networks, which make them unsuitable for use in an SDWN architecture. Some scholars have modeled the routing optimization problem as either a linear programming (LP) problem or a nonlinear programming (NLP) problem \cite{b7} and proposed the use of classical routing optimization algorithms to find the optimal routing strategy. However, as the network scale increases, it becomes difficult for such an algorithm to dynamically adjust the routing and forwarding strategy based on the changing network topology and link information. Some researchers have proposed improving the effectiveness and stability of routing algorithms by performing predictive processing on the available network traffic information before routing optimization \cite{b8}. By doing so, accurate traffic information can be obtained promptly to help the routing algorithm make precise path decisions. With the development of artificial intelligence technology, many researchers have begun to apply related methods to routing optimization problems to improve the performance of large-scale and complex dynamic networks and simultaneously optimize multiple network links. In particular, deep reinforcement learning (DRL) \cite{b9} is useful for solving goal-oriented learning and decision-making problems and thus can help routing algorithms flexibly adapt to complex, dynamic changes in wireless networks to optimize the allocation of wireless network resources. For the wireless network routing problem discussed in this article, DRL utilizes deep neural networks that can handle high-dimensional and complex state and action spaces, better adapt to a dynamic network environment, and adaptively adjust the network routing strategy to maximize the overall network efficiency. In previous work, DRL has already achieved some significant results in routing optimization \cite{b9}$-$\cite{b12}. 
 
The article proposes an intelligent SDWN routing algorithm, DRL-PPONSA, based on proximal policy optimization (PPO) DRL and network situational awareness (NSA) \cite{b10}. First, this algorithm comprehensively considers the traffic information on the links in the wireless network. To this end, the data plane and control plane of the SDWN architecture and a graph convolutional network–gated recurrent unit (GCN-GRU) \cite{b11} prediction mechanism are designed as three key components of the NSA system. The control plane is responsible for data perception and acquisition, while the GCN-GRU mechanism is applied for traffic prediction, thereby achieving awareness of the global network state. Next, traffic matrices representing the predicted remaining bandwidth, link delay, packet loss rate, packet error rate, and wireless node distance as well as the network topology are taken as the description of the environment for DRL, while next-hop adjacent nodes are viewed as possible actions. By utilizing the link information in the network environment and employing various reward and punishment mechanisms, a reward function is designed to guide a DRL agent to learn a strategy seeking the highest reward, ultimately achieving optimal intelligent routing decisions.
 
The contributions of this paper are summarized as follows:
 
 \begin{enumerate}
 	\item  Compared to the relatively simplistic collection of network status information previously described in the literature, this paper comprehensively considers multiple types of status information, such as transmission delay, propagation delay, network jitter, remaining bandwidth, packet loss rate, packet error rate, and distance between access points, and uses them as optimization objectives. This approach can more comprehensively satisfy QoS requirements, laying a foundation for efficient and flexible data transmission and effectively reducing network congestion.
 	\item Compared to traditional routing algorithms that make routing decisions based on network information measured at fixed intervals, this paper presents an NSA mechanism using a GCN-GRU prediction model based on an SDWN architecture. This mechanism comprehensively considers the spatiotemporal characteristics of wireless network traffic, using current and historical traffic status data to predict the future traffic situation. This approach can fully utilize the hidden status information in a wireless network to improve the reliability and stability of intelligent routing algorithms.
 	\item In contrast to schemes that use k-paths action design and a single reward function, the interaction between an intelligent agent and the network environment is enhanced in this paper by designing a novel action space for DRL that consists of the next-hop adjacent nodes. This makes the routing algorithm more intelligent and adaptable to dynamic and complex network changes. Additionally, different reward functions are designed based on the link information and reward and punishment mechanisms to reduce the risk of the intelligent agent becoming trapped in a local optimum, thereby improving the robustness and effectiveness of the intelligent routing algorithm.
 	\item In contrast to offline value-based reinforcement learning methods, this paper uses an online DRL method based on PPO to address the problems of slow convergence and high storage space occupation for offline experience caching. During the gradient update process, to solve the problem of low sample effectiveness due to mismatch between the probability distribution of the sampled data and the current agent policy as well as the problem of network gradient explosion, importance sampling and gradient clipping are used instead of prioritized experience replay and the Kullback–Leibler (KL) divergence. This approach improves the convergence speed and stability of the intelligent algorithms.
 \end{enumerate}

The remainder of this article is organized as follows: The second section introduces mainstream prediction models and routing methods along with their associated shortcomings. The third section introduces the proposed intelligent routing architecture based on NSA under the SDWN architecture. The fourth section introduces the environment design for the proposed DRL-PPONSA algorithm and the details of algorithm implementation for intelligent routing. The fifth section reports experiments conducted to verify the stability and reliability of the algorithm proposed in this paper. The sixth section is the conclusion, which summarizes the challenges faced and proposes further research directions.

\section{Related Work}
This section introduces commonly used routing optimization methods and traffic prediction methods used to improve the reliability and stability of routing algorithms. The mainstream routing optimization methods are divided into classical optimization algorithms and intelligent optimization methods, while the traffic prediction methods are also divided into two main categories: linear prediction and nonlinear prediction.

\subsection{Routing Optimization Methods}
Routing optimization is a key technology for improving network performance, providing important guarantees for network scalability, stability, and security.  

\subsubsection{Classic routing optimization methods}
 This section introduces classical route optimization algorithms based on heuristic algorithms. Common algorithms of this kind include simulated annealing (SA), genetic algorithms (GAs), ant colony optimization (ACO), and artificial neural networks (ANNs). They are bioinspired optimization algorithms proposed based on the behavior of natural organisms. Yao et al. \cite{b14} used an improved GA to solve the routing problem for a wireless sensor network (WSN) by comprehensively considering the distance between adjacent nodes, energy consumption, communication delay, and other factors to select appropriate routes. In addition, Ke et al. \cite{b15} proposed using an artificial bee colony algorithm to solve the routing and forwarding problem for sensor nodes in dynamic SDN networks. By detecting the traffic status information of the wireless sensors through SDN technology, the routing strategy can be dynamically adjusted, improving the energy and data allocation situation of the sensor nodes. Yong et al. \cite{b16} proposed a method of combining an SA algorithm and a GA to find the optimal routing strategy in an SDN network to overcome the challenges of wide distribution, high latency, and large data volume in a power grid system. First, the SA algorithm was used to search for feasible solutions to the problem, considerably reducing the computational overhead. Then, the GA was used to find the optimal solution for routing deployment. Truong et al. \cite{b17} proposed a heuristic traffic engineering algorithm for use in SDN networks with multipath forwarding and inter-path traffic switching. First, the k-paths approach was applied to find the initial lowest cost path selection, and then the heuristic algorithm was executed to make routing decisions in accordance with the path load and traffic conditions to achieve the optimal multipath forwarding strategy. Shokouhifar et al. \cite{b18} proposed using a fuzzy heuristic-based ACO algorithm to adaptively adjust the routing placement in the network, taking into account delay and reliability factors. Sowmya et al. \cite{b19} used a long short-term memory (LSTM) network to dynamically predict the flow and path information in an SDN network, then used ANN modeling to predict the paths of data packets and executed routing in accordance with the perceived network state information.

Although classical optimization algorithms can solve shortest-path optimization problems to some extent, these algorithms typically provide only suboptimal solutions and often optimize only a single objective. Subtle changes in the network environment can cause significant fluctuations and errors in these algorithms, leading to potential scalability issues and impacting network performance \cite{b8}. Therefore, classical routing optimization algorithms are not suitable for large-scale, dynamic, and complex SDWN environments.

\subsubsection{Intelligent routing optimization methods}
With the continuous development of science and technology, traditional classical routing algorithms are increasingly unable to meet complex and dynamic network demands. In contrast, intelligent optimization algorithms can better adapt to complex and high-dimensional dynamic network environments. Therefore, many researchers have studied the application of intelligent algorithms to routing optimization problems and achieved good results. Xiang et al. \cite{b20} proposed an energy-aware routing algorithm (EAMWPA) for use in the SDWN context. This algorithm divides the energy of the network nodes into several levels and then horizontally adjusts the weight information of the links based on the energy of all nodes, ultimately achieving a balance of energy among nodes and extending the network's lifecycle. Duong et al. \cite{b21} proposed a machine learning method (IRSML) for route optimization in SDWN networks to optimize multiple targets at the same time. First, supervised learning was used to predict the performance indicators of links. The main indicators considered were the minimum end-to-end delay, transmission quality, and packet blocking probability, which were used to characterize the network states; the next hop was considered for action selection, and Q-learning reinforcement learning was used to determine the goals and constraints of the routes. Smita et al. \cite{b22} proposed a routing algorithm combining Q-learning algorithm and a feedforward neural network to address the adaptive routing problem in wireless mesh networks. Bo et al. \cite{b23} presented the use of GNN and deep reinforcement learning methods to address the poor generalization of traditional routing methods in dynamically changing network topologies. Murtaza et al. \cite{b4} proposed a machine learning-assisted routing algorithm based on SDWN in which the routing strategy was dynamically adjusted based on the historical network parameters (delay, bandwidth, signal-to-noise ratio, etc.) of mobile nodes.

Casas et al. \cite{b24} proposed a deep Q-network (DQN)-based DRL scheme to generate dynamic traffic changes for an SDN-based intelligent routing strategy, considering that reinforcement learning cannot adapt well to high-dimensional state and action spaces. The source–destination node pairs were used as the state space, and the k-paths schemes from the source to destination nodes were used as the action space. Huang et al. \cite{b8} used a GRU model to predict the traffic information of an SDN network and then constructed a traffic matrix using parameters such as the residual bandwidth, delay, and packet loss rate as the state space for a Dueling DQN-based DRL algorithm. Using the k-paths approach for intelligent action selection, they dynamically searched for the best routing strategy. Zhao et al. \cite{b3} used the DRL-M4MR method to optimize intelligent multicast routing in the SDN network of a New York data center. They constructed a state matrix based on link information such as link bandwidth, delay, and packet loss rate and considered each edge of a link as a possible next hop in the action space, thus intelligently constructing an optimal multicast tree and achieving multicast flow table deployment. Liu et al. \cite{b25} proposed a DRL-R model using the DQN and deep deterministic policy gradient (DDPG) algorithms to quantify cache and bandwidth resource allocation, thereby reducing unnecessary network delay. The convergence effect of this model was better than that of a DQN model alone. The algorithm was designed to use QoS-related parameters to characterize the state space and adopted the k-paths scheme to generate the preselected actions of the intelligent agent. Zhang et al. \cite{b26} proposed a solution combining message passing neural networks with DRL to address the challenges in routing decision making brought by dynamic wireless network topologies. Guo et al. \cite{b27} proposed a DDPG-based DRL approach for the switch deployment problem based on multiple traffic matrices. They first clustered historical traffic data and obtained a representative traffic matrix for training. Then, they used the state matrix as the state space of the intelligent algorithm to find the optimal positions for SDN switch placement. Sun et al. \cite{b28} used a GRU model to extract temporal features from traffic and dynamically adjusted critical link weight information through an actor–critic DRL algorithm. They used the traffic distribution on each link as the state information of the intelligent algorithm and added a random value based on the k-paths method to improve the anti-interference ability of the intelligent agent, thus expanding the search range in the intelligent agent's action space to achieve the optimal routing strategy.

In the abovementioned literature, classical machine learning algorithms were used to optimize network routing and improve network performance. Although these approaches can optimize multiple target parameters simultaneously, doing so requires a large amount of labeled data for training, and obtaining corresponding label information from dynamic and complex network topologies is extremely difficult and highly dependent on the accuracy of the data. If the obtained dataset is inaccurate, the system will not be able to learn a good network model to predict the optimal routing strategy. Therefore, many researchers have begun to focus on reinforcement learning as an optimization method for intelligent routing. Compared with traditional routing algorithms and classical optimization algorithms, reinforcement learning methods can greatly improve network performance and have significant advantages in adjusting routing strategies in dynamically changing networks in a timely manner. Therefore, reinforcement learning is also a feasible technology for solving routing optimization problems in large-scale and complex dynamic wireless networks. However, most of the related literature to date has focused on intelligent routing problems in wired SDN environments, and there has been little in-depth discussion of wireless network routing.

\subsection{Traffic Prediction Methods}
Network traffic information is an important feedback indicator of network load. Accurately and reliably predicting network traffic information can help avoid delays caused by network congestion and assist routing algorithms in achieving fault-tolerant processing, thus improving their stability and reliability. Typically, the algorithms used for network state prediction can be divided into linear prediction methods and nonlinear prediction methods.

\subsubsection{Linear prediction methods}
The most commonly used linear prediction models include autoregressive moving average (ARMA) models, autoregressive integrated moving average (ARIMA) models, Markov models, and the Holt–Winters algorithm. Tian et al.\cite{b29} proposed modeling the frequency component of network traffic with an ARMA model to predict network traffic in an SDN environment. Alghamdi et al. \cite{b30} proposed using an ARIMA model to predict traffic flow in the state of California in the United States. Tran et al.\cite{b31} proposed an improved algorithm based on Holt–Winters exponential smoothing to predict abnormal data in cellular traffic environments. However, traditional linear models have simple structures and have difficulty accurately perceiving fast-changing network traffic features, resulting in poor prediction results as well as weak model adaptability and generalization ability. The traffic information in wireless networks includes not only temporal features but also spatial features \cite{b21}. Therefore, using linear models to predict wireless traffic has certain limitations.

\subsubsection{Nonlinear prediction methods}
For complex and diverse types of network data, accurate prediction can be achieved through nonlinear modeling; for example, nonlinear models such as neural networks and fuzzy logic models can be used to predict network traffic. Casado-Vara et al. \cite{b32} proposed using an LSTM recurrent neural network trained through distributed asynchronous training to predict web traffic time series, fully exploring the hidden temporal features of the network traffic and improving the accuracy of prediction. Hu et al. \cite{b33} proposed using an improved variant of an LSTM recurrent neural network model in which the forget gate and input gate are combined into an update gate to predict data information in network traffic. Bi et al. \cite{b34} proposed using a Savitzky–Golay (SG) filter to preprocess the noise information in the original traffic data, using a temporal convolutional network (TCN) to extract the short-term features of the sequence, and using an LSTM network to capture the long-term dependencies in the data, effectively capturing the nonlinear features of the network sequence and improving the prediction accuracy. Huang et al.\cite{b8} proposed using a GRU model instead of an LSTM model to predict traffic matrix information in the SDN context, extracting hidden traffic information from the network to construct a corresponding predicted traffic matrix, and using the predicted traffic matrix to train an intelligent agent, thereby improving the reliability of the routing algorithm.

Although related studies have addressed the issues of low linear prediction accuracy and weak generalization ability and have achieved good prediction results, they have only considered the temporal features of network traffic and have not taken spatial features into account. For wireless network traffic with spatiotemporal characteristics, the prediction effect is not ideal. Therefore, Yuan et al. \cite{b35} proposed a method combining a 3D convolutional neural network (3D-CNN) and an LSTM network to predict traffic information in wireless networks. Pan et al. \cite{b36} fully considered the complex temporal and spatial dependencies among network traffic and proposed a prediction model combining the GCN and GRU techniques to predict network traffic. The experimental results showed that this model has long-term prediction capabilities and is applicable for traffic data with spatiotemporal characteristics, and the prediction effect meets expectations. Therefore, in this paper, a network traffic prediction scheme is designed that also combines the GCN and GRU approaches for the accurate prediction of wireless traffic with spatiotemporal features. The integration of this GCN-GRU prediction algorithm into the intelligent routing algorithm proposed in this paper is an important means of improving its reliability and effectiveness.

\section{THE DESIGN OF THE PROPOSED INTELLIGENT ROUTING ARCHITECTURE BASED ON SDWN FOR NSA}
\begin{figure}[t!]
	\centering
	\includegraphics[width=\linewidth]{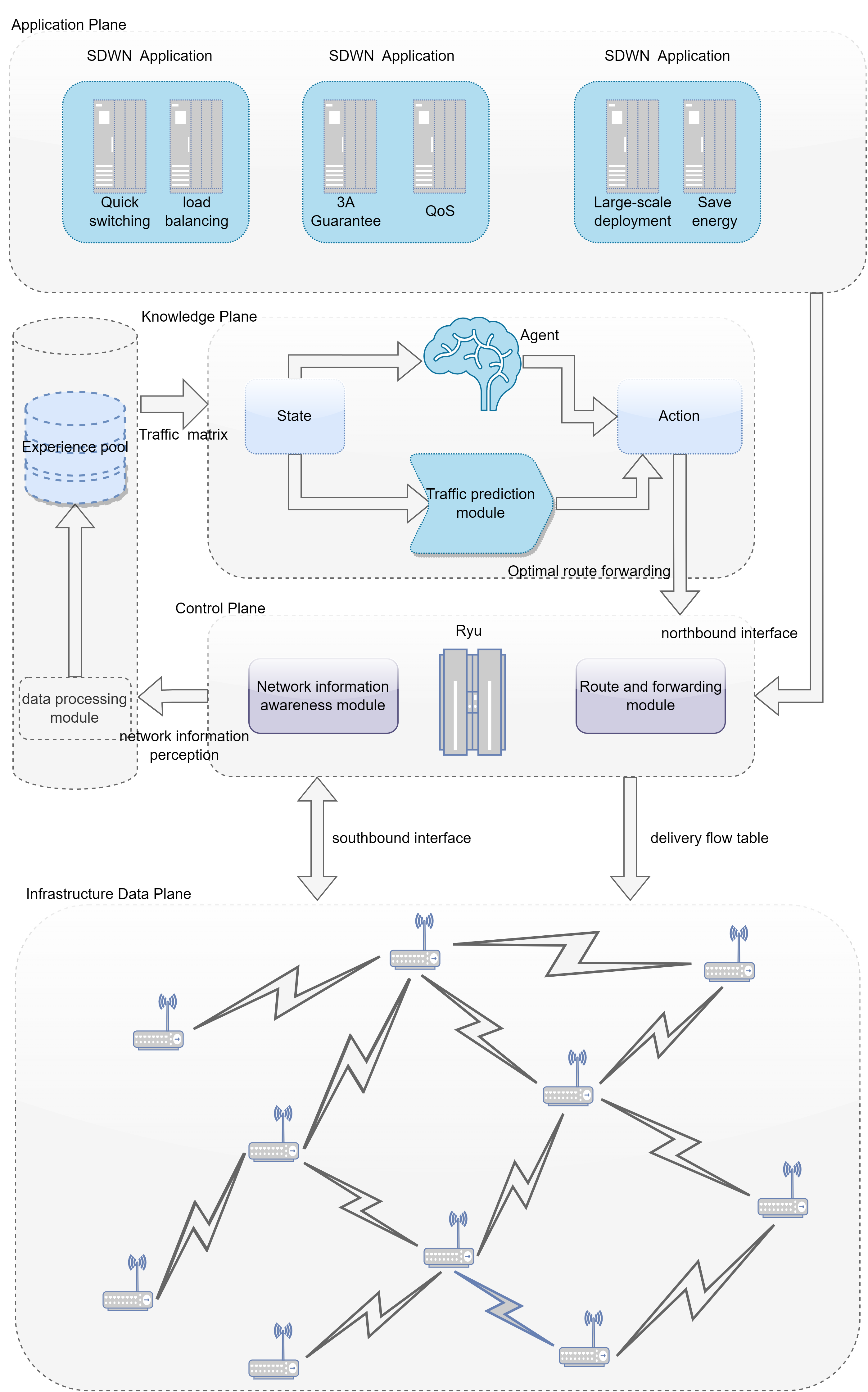}
	\caption{\textbf{SDWN intelligent routing structure.}}
	\label{fig1}
\end{figure}

In this section, we introduce the proposed intelligent routing architecture designed in this paper based on SDWN for NSA. The overall architecture of the intelligent routing system is composed of an infrastructure data plane, a logical control plane, a knowledge plane, and an application plane, as shown in Fig. \ref{fig1}. The structure of each plane is presented in detail below.

\subsection{Infrastructure Data Plane}
This plane primarily consists of wireless network devices such as APs and stations (STAs). This plane is responsible for packet lookup and forwarding as well as packet parsing and flow table matching. The devices respond to requests sent by the logical control plane via the southbound interface of the OpenFlow protocol.

\subsection{Logical Control Plane}
SDWN is a method of utilizing SDN technology to achieve centralized control in wireless networks. Accordingly, rules defined by specialized programs, usually referred to as controllers, determine the network behavior. In SDWN, the wireless control plane and data plane are decoupled and separated, thereby simplifying the wireless access devices and forwarding devices and allowing the network to run in accordance with the rules of logical centralized control plane scheduling. First, the SDWN controller in the control plane periodically collects information from the data plane, such as the network topology, link bandwidths, link round-trip delays, link packet loss rates, and distances between wireless APs, through its southbound interface.

The main functions of the controller include link discovery, topology management, policy formulation, and flow table distribution. Thereby, in link discovery and topology management, the controller utilizes an uplink channel to uniformly monitor and aggregate the information reported by the underlying switch devices. Policy formulation and flow table distribution are achieved through the use of downstream channels to implement unified control over the network devices.

The purpose of link discovery in SDWN is to obtain global network information that serves as the foundation for network address learning, VLAN management, and routing/forwarding \cite{b37}. Unlike traditional network link discovery methods, link discovery in SDWN networks is accomplished uniformly through the link layer discovery protocol (LLDP) by the controller. The objective of topology management is to collect and monitor information from the SDWN switches in the network in order to provide feedback on the working status of the switches and the connection status of the links \cite{b38}. The controller regularly sends LLDP packets through packet-out messages to the connected SDWN APs and senses the information of these wireless APs based on the feedback obtained from packet-in messages, thereby detecting the working status of the wireless switches and updating its representation of the network topology.

Unlike in most other literature that considers only a single type of network state information, the network state information addressed in this article includes the remaining link bandwidth $bw_{free(ij)}$, the used link bandwidth  $bw_{use(ij)}$, the link delay $delay_{ij}$, the packet error rate $pkt_{err(ij)}$ on a link, the number of dropped packets $pkt_{drop(ij)}$ on a link, the distance $distance_{ij}$ between wireless APs, and the packet loss rate  $loss_{ij}$ on a link. The controller operates with a cycle time of $ t $ seconds and sends port information query requests to the data plane. The data plane responds with a reply that contains the statistics for each port, including the numbers of bytes transmitted  $tx_b$ and received $rx_b$, the number of packets transmitted $tx_p$ and received $rx_p$, as well as the effective time $t_d$. The difference between two consecutive sets of statistics represents the amount of bandwidth used during that period. For a link $e_{(i,j)}$ , the maximum used bandwidth $bw_{use(ij)}$ is determined by taking the maximum value of the used bandwidth between the two ports ($i$ and $j$) on the two switches connected by the link. The remaining bandwidth $bw_{free(ij)}$ on the link is equal to the difference between the maximum capacity of the link, $bw_{capmax}$, and the used bandwidth, $bw_{use(ij)}$, as shown in Equation \eqref{eq1}. Equation \eqref{eq2} is used to calculate the maximum used bandwidth of the link, where $\left ( i,j \right ) \in \left ( 1,2,3,…,n \right ) ,i\ne j$, the $n$ represent the number of AP switches.
\begin{equation}
b{w_{free\left( {ij} \right)}} = \left| {b{w_{capmax}} - b{w_{use\left( {ij} \right)}}} \right|\label{eq1}\end{equation}
\begin{equation}
b{w_{use\left( {ij} \right)}} = \left| {\frac{{\left| {\left( {t{x_{bj}} + r{x_{bj}}} \right) - \left( {t{x_{bi}} + r{x_{bi}}} \right)} \right|}}{{{t_{dj}} - {t_{di}}}}} \right|\label{eq2}\end{equation}

The round-trip delay calculated for a link is detected by the controller's built-in Switches module, and the timestamp of LLDP [38] data transmission is obtained. The controller sends an echo request message with a timestamp to the AP switches, then parses the echo reply message returned by each switch and subtracts the transmission time parsed from the data from the current time to obtain the echo round-trip delays $T_{echo\_api}  $ and $T_{echo\_apj} $ between the controller and the wireless switches. The delay between the wireless switches is then calculated from the LLDP message receiving time minus the message sending time for each switch, denoted by  $T_{lldp\_api} $ and $T_{lldp\_apj} $. Accordingly, the $delay_{ij}$ calculation for link $e_{(i,j)}$ is as show in \eqref{eq3}.
\begin{equation}
dela{y_{ij}} = \frac{{\left( {{T_{lldp\_api}} + {T_{lldp\_apj}} - {T_{echo\_api}} - {T_{echo\_api}}} \right)}}{2}.\label{eq3}\end{equation}

Equation \eqref{eq4} represents the maximum packet loss rate calculated for link $e_{(i,j)}$ from switch port $i$ to $j$ and from switch port $j$ to $i$ . Equation \eqref{eq5} represents the packet error rate of link $e_{(i,j)}$ in both directions, from switch port $i$ to $j$  and from switch port $j$ to $i$. In Equation \eqref{eq6}, the number of dropped packets for link $e_{(i,j)}$ represents the number of packets lost in both directions, from switch port $i$ to $j$ and from switch port $j$ to $i$.
\begin{equation}
los{s_{ij}} = \max \left( {1 - \frac{{r{x_{pj}}}}{{t{x_{pi}}}},1 - \frac{{r{x_{pi}}}}{{t{x_{pj}}}}} \right) \cdot 100\% \label{eq4}\end{equation}

\begin{equation}
pk{t_{err\left( {ij} \right)}} = \left| {\frac{{r{x_{bi}} - t{x_{bj}}}}{{r{x_{bi}}}}} \right| \cdot 100\%  \label{eq5}\end{equation}

\begin{equation}
pk{t_{drop\left( {ij} \right) = }}\left| {r{x_{pi}} - t{x_{pj}}} \right| \label{eq6}\end{equation}
where $tx_{bi}$ represents the number of bytes sent from port $i$,  $rx_{bi}$ represents the number of bytes received at port $i$. $tx_{pi}$ represents the number of packets sent from port $i$, $rx_{pi}$ represents the number of packets received at port $i$, $\left ( i,j \right ) \in \left ( 1,2,3,…,n \right ) ,i\ne j$, the $n$ represent the number of AP switches.

In actual physical scenarios, the distance traveled corresponds to the energy consumed for wireless transmission; therefore, the algorithm in this article also considers the distance between APs. The Equation \eqref{eq7} gives the distance $distance_{ij}$ between two APs, which represents the distance in physical space between the wireless AP switches and is calculated using three-dimensional coordinates.
\begin{equation}
distanc{e_{ij}} = \sqrt {{{\left( {{x_i} - {x_j}} \right)}^2} + {{\left( {{y_i} - {y_j}} \right)}^2} + {{\left( {{z_i} - {z_j}} \right)}^2}} \label{eq7}\end{equation}
where $\left(x_i,y_i,z_i\right)$ represents the spatial three-dimensional coordinates between of the i-th switch.

The controller formulates forwarding strategies and generates corresponding flow table entries based on the transmission requirements at the different levels. A designed DRL mechanism in the knowledge plane learns the optimal routing strategy based on current network state information, and issues flow tables to generate optimized SDWN forwarding routes.

\subsection{Knowledge Plane}
\begin{figure}[t!]
	\centering
	\includegraphics[width=\linewidth]{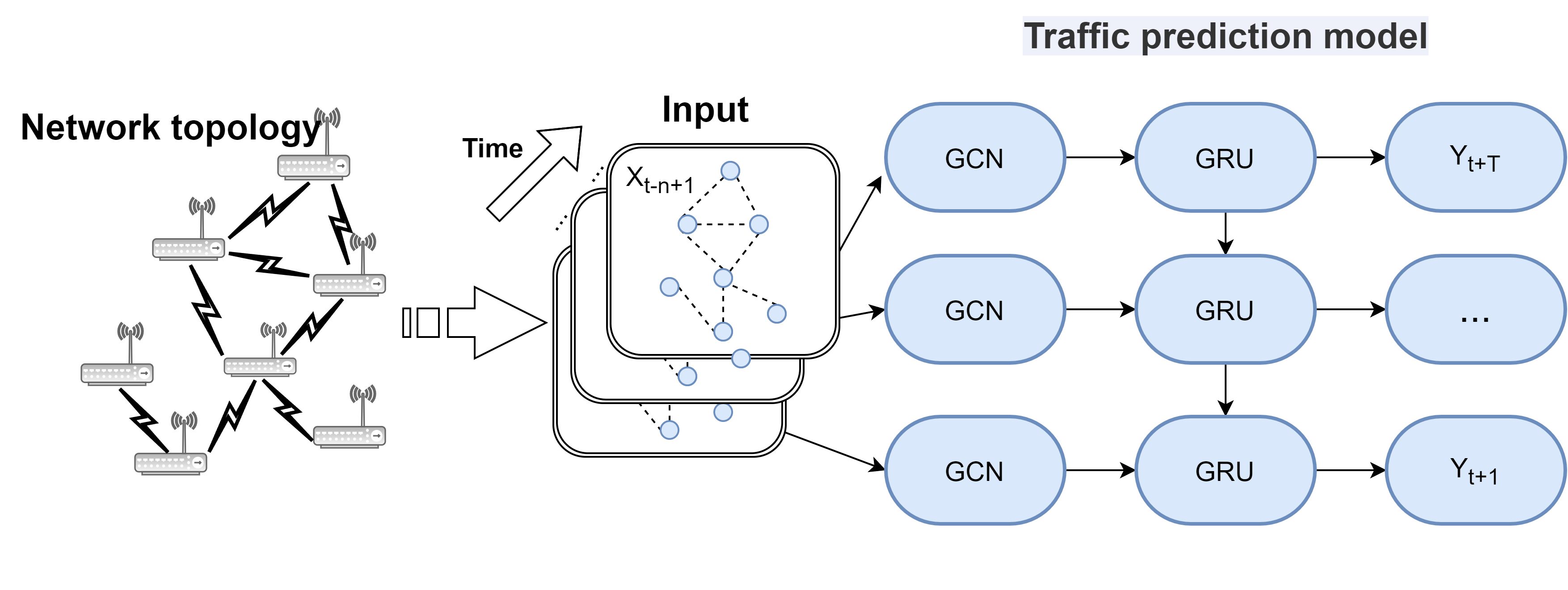}
	\caption{\textbf{Traffic prediction module.}}
	\label{fig2}
\end{figure}

The knowledge plane designed in this article applies a combination of reinforcement learning based on policy gradients and traffic prediction. Loading the knowledge plane into the SDWN framework endows the controller with greater intelligence when making policy decisions. Model training in the knowledge plane requires data obtained from the information pool in advance and network link information converted into a traffic matrix, which is predicted by a traffic prediction module. The traffic prediction module consists of two parts: a GCN and a GRU network. As shown in Fig. \ref{fig2}, the topology information of the SDWN network is first obtained, and the Ryu controller obtains the link information and converts it into graph data with weights, which are treated as time series data at t time points. Then, these time series data at t historical time points are used as input for prediction. First, the spatial features of the wireless network topology are obtained through the GCN. Second, the time series data with spatial features are input into the GRU network for the extraction of temporal features. Finally, a fully connected layer is used to filter and output the predicted results. The obtained predicted traffic data are synchronized into a reinforcement learning network for training, and an intelligent agent optimizes the strategy to be executed based on the currently obtained network state information, with the maximum reward value as the target, dynamically adjusting the optimal routing path until the model converges and stabilizes. Finally, the trained model is deployed to the SDWN controller to obtain the optimal routing strategy. Data flow table forwarding is completed by responding to the data plane through the southbound interface.

\subsection{Application Plane}
The application plane is a system that manages all business-related applications and has a programmable API. Controllers interact with the application plane through their northbound interfaces to enable the development and deployment of various network applications, such as load balancing, fast switching, and interference management.

\subsection{Design of the Proposed NSA System Architecture Based on SDWN}
To enable the proposed intelligent routing algorithm to better optimize the network performance with enhanced stability and effectiveness, this article presents the design of an NSA architecture based on SDWN. This NSA architecture is used to monitor and obtain information on the traffic and network topology in the wireless network, predict future trends in network traffic, analyze network conditions from a global perspective, and provide accurate data support for routing. Meanwhile, DRL methods are used to search for the optimal wireless routing/forwarding paths to improve network bandwidth utilization and reduce the network delay, thereby ensuring the communication quality of the entire network.

\begin{figure}[t!]
	\centering
	\includegraphics[width=\linewidth]{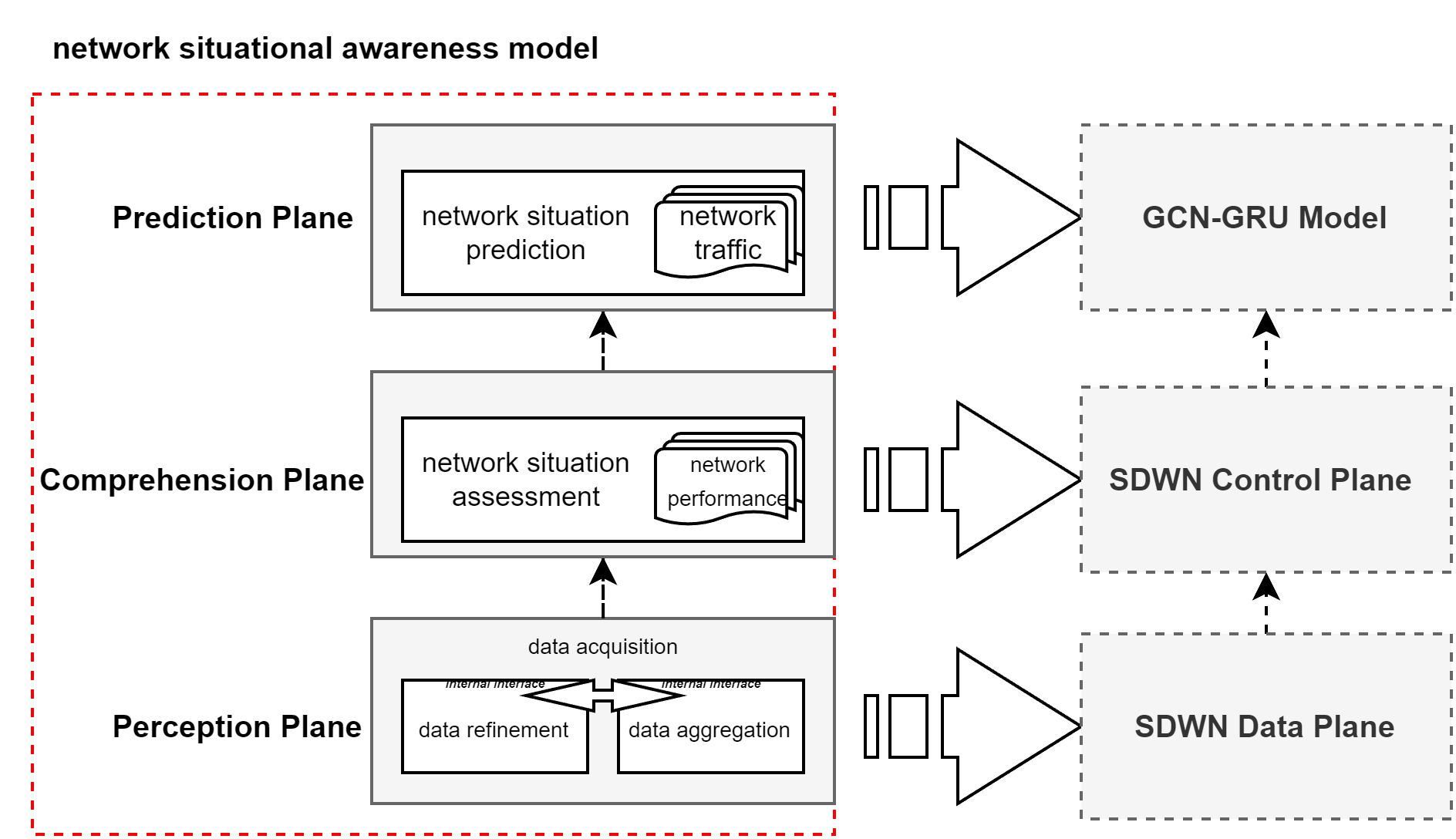}
	\caption{\textbf{Architecture of network situation awareness system.}}
	\label{fig3}
\end{figure}

Situational awareness refers to the ability to perceive environmental factors and events in time or space as well as predict their future states \cite{b39} by processing existing information and seeking an optimal solution on a specific timeline. The application of NSA in intelligent routing mainly involves perceiving and analyzing real-time data from the wireless network to optimize and schedule the network topology and wireless network traffic, thereby improving the network throughput and response speed and providing important guarantees of network security. As shown in Fig. \ref{fig3}, the NSA system architecture proposed in this article consists of three layers: a perception layer, a comprehension layer, and a prediction layer. Among them, the perception layer mainly collects and processes data information, corresponding to the data plane in SDWN. The comprehension layer corresponds to the control plane in the SDWN framework; in this layer, the Ryu controller comprehends and projects data information from the perception layer and evaluates the current situation of the network environment. As the highest level of situational awareness, the prediction layer is responsible for perceiving and understanding various elements of the network environment and predicting future traffic trends; this layer corresponds to the traffic prediction model in this article.

\section{INTELLIGENT SDWN ROUTING ALGORITHM BASED ON POLICY OPTIMIZATION, DRL-PPONSA}
\begin{figure}[t!]
	\centering
	\includegraphics[width=\linewidth]{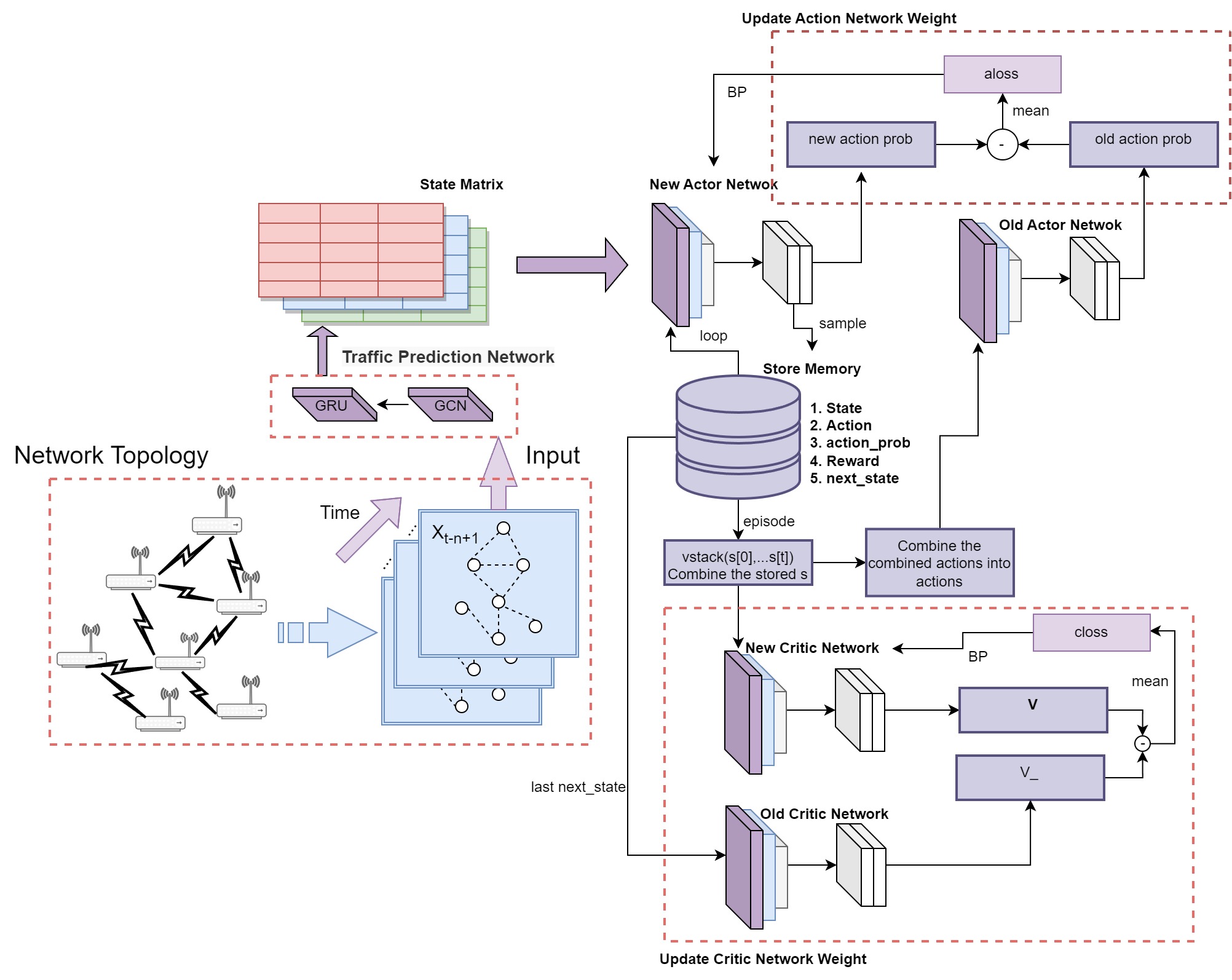}
	\caption{\textbf{DRL-PPONSA algorithm framework.}}
	\label{fig4}
\end{figure}

In the infrastructure data plane, the wireless network topology is abstracted as an undirected graph with relevant link parameter weights, represented by $G=\left(G,E,W\right)$. Where $V$ is the set of wireless APs, $E$ is the set of links between nodes, $W$ is the set of link parameter information, and each side $ e_{ij} $ represents a link, $e_{ij}\in E$. Accordingly, the intelligent routing algorithm in this article is designed based on an undirected graph. A diagram of the overall algorithm is shown in Fig. \ref{fig4}. First, the link information of the network is obtained based on the SDWN topology, including parameters such as bandwidth, delay, and packet loss rate, and is transformed into a multidimensional matrix. Then, the predicted traffic matrix is obtained through the traffic prediction model, and finally, the predicted traffic matrix is provided as input to the DRL algorithm for training. The DRL algorithm is composed of two main networks: an actor network and a critic network. The former optimizes a policy to maximize the expected returns, while the latter evaluates the value of a given policy in its current state. These two networks collaborate to achieve efficient policy optimization. This approach based on an AC network architecture can ensure that the intelligent agent always makes decisions in the direction of accumulating the maximum reward while dynamically optimizing parameters such as network delay, jitter, bandwidth, and throughput rate to achieve real-time control of the current network, thereby effectively reducing the network load \cite{b40}. This section introduces the design of the network traffic prediction algorithm and the DRL algorithm and finally describes the DRL-PPONSA algorithm framework proposed in this paper.

\subsection{Design of the Network Traffic Prediction Algorithm}
The network traffic components in an SDWN environment are complex, and implementing a single method for traffic tracking and statistics can be expensive and may not produce satisfactory results. Previous research on SDWN network traffic has mainly focused on the traffic matrix (TM) \cite{b41}. In this article, we propose a new approach that combines the GCN and GRU techniques. This approach takes into account both the spatial and temporal characteristics of the data to improve the ability of an intelligent agent to accurately predict the future TM.

As shown in Algorithm \ref{alg:algorithm1}. we take the converted the converted TM $X_{1,2,3…,t}$, the adjacency matrix $A$ of the graph, the learning rate $\alpha $, the number of training episodes $\mathcal{M}$, the batch size $\mathcal{K}$, and the $ L_2 $ regular term coefficient $\lambda $ as inputs. On this basis, the algorithm will output the predicted TM $Y_{1,2,3…,t}$. The first and second lines randomly initialize the network weights $\theta$ and standardize the input $X_{1,2,3…,t}$. The third line calculates the degree matrix parameters $ D $ used in the prediction network convolution from the adjacency matrix $ A $. The fourth to eighth lines describe the iterative network training process. The GCN-GRU network generates the hidden layer parameters $H$. Finally, the output is produced by filtering the prediction results through a fully connected layer. During training, the weight parameters of the network are updated based on the minimum quantized true value and the square difference of the traffic prediction results. The predicted traffic matrix output by the GCN-GRU traffic prediction algorithm will serve as the input to the reinforcement learning algorithm.

\begin{algorithm}[htp]
   \small
    \caption{GCN-GRU Network Traffic State Prediction Algorithm}
	\label{alg:algorithm1}
	\begin{algorithmic}[1]
		\Require  traffic matrix data:$X_{1,2,3,...,t}$, graph adjacency matrix:$A$, learning rate:$\alpha$, training episodes:$\mathcal{M}$, batch size:$\mathcal{K}$, $ L_2 $ regularization coefficient:$\lambda$.
		\Ensure
		 Predicted traffic matrix data:$Y_{t+1,t+2,t+3,...,t+T}$.
		\State \textbf{Initialize} GCN-GRU network with random weight:$\theta$.
     	\State \textbf{Input} data for standardization.
     	\State \textbf{Obtain} adjacency matrix A of SDWN network topology and calculate $\widehat{\mathrm {D}} ^{-1/2}\widehat{\mathrm {A}}\widehat{\mathrm {D}} ^{-1/2}$
 	    \For{$episode=1 \gets \mathcal{M} $}
	     \State Generate implicit H according to GCN-GRU network
	     \hspace*{2.5em} $H_{t+1,...,t+T}=GCGRU(X_{t-n+1},\widehat{\mathrm {D}} ^{-1/2}\widehat{\mathrm {A}}\widehat{\mathrm {D}} ^{-1/2})$ 
	     \State Calculate the final prediction result based on FC network 
  	     \hspace*{2.5em} $Y_{t+1,...,t+T}=FC(X_{t+1,...,t+T})$ 
	     \State  Calculate loss function: $Loss=\frac{1}{\mathcal{K}} \sum_{i=1}^{\mathcal{K}} \left (  Y_{t}-\widehat{Y_{t}}  \right )^{2}+ \hspace*{2.5em} \frac{1}{2} \lambda  \left \| \theta \right \| _{2}^{2}  \to  0 $, update model weights $\theta =\widehat{\theta } $
	    \EndFor
		
	\end{algorithmic}
\end{algorithm}

\subsection{Design of DRL-PPO Intelligent Routing Algorithm}
\begin{figure}[t!]
	\centering
	\includegraphics[width=\linewidth]{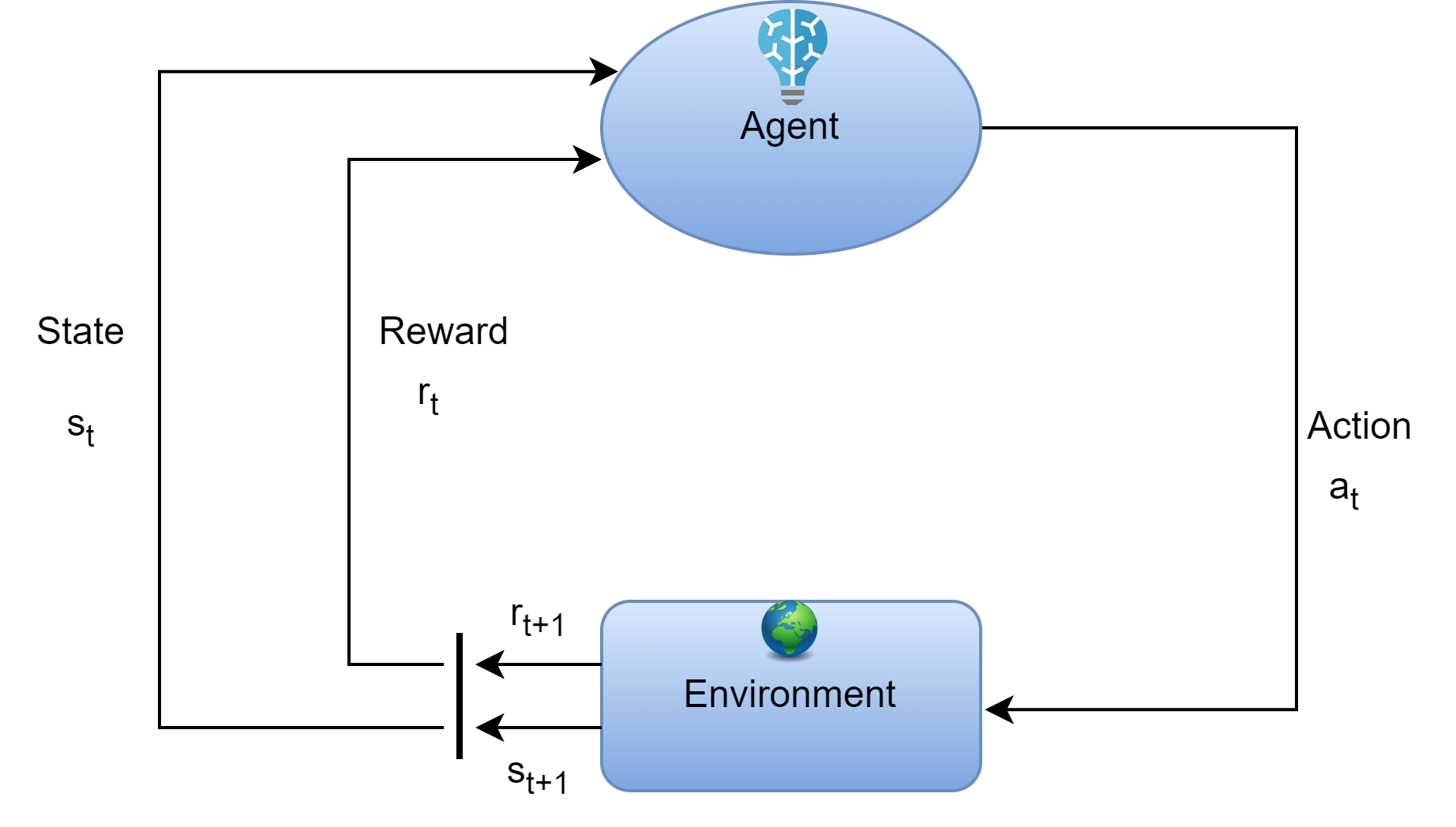}
	\caption{\textbf{The interaction process between agent and environment in reinforcement Learning.}}
	\label{fig5}
\end{figure}

The most important task when solving wireless network routing optimization problems through DRL algorithms is to design a state space, a reward function, and an action space for decision-making that match the current environment. The state space describes the current state that a DRL agent can observe from the environment. The agent perceives the current network state information, and its learning direction in updating its policy is then guided by the reward function $ R $. After the agent executes the action $A_t$ in state $S_t$, the current state transitions to the next state $S_{t+1}$, and $ A_t $ receives the corresponding reward value $R_{t+1}$ according to $ S_{t+1} $. When the agent obtains the maximum reward value, the path chosen by the agent is the optimal path. The corresponding interaction process is shown in Fig. \ref{fig5}.

In the following, we provide a detailed introduction to the design of the state space, action space, and reward function in the DRL algorithm proposed in this article as well as the importance sampling method, gradient clipping optimization scheme, and PPO update strategy used in the PPO algorithm \cite{b42}.

\subsubsection{State space $\mathcal{S}$}
\begin{figure}[t!]
	\centering
	\includegraphics[width=\linewidth]{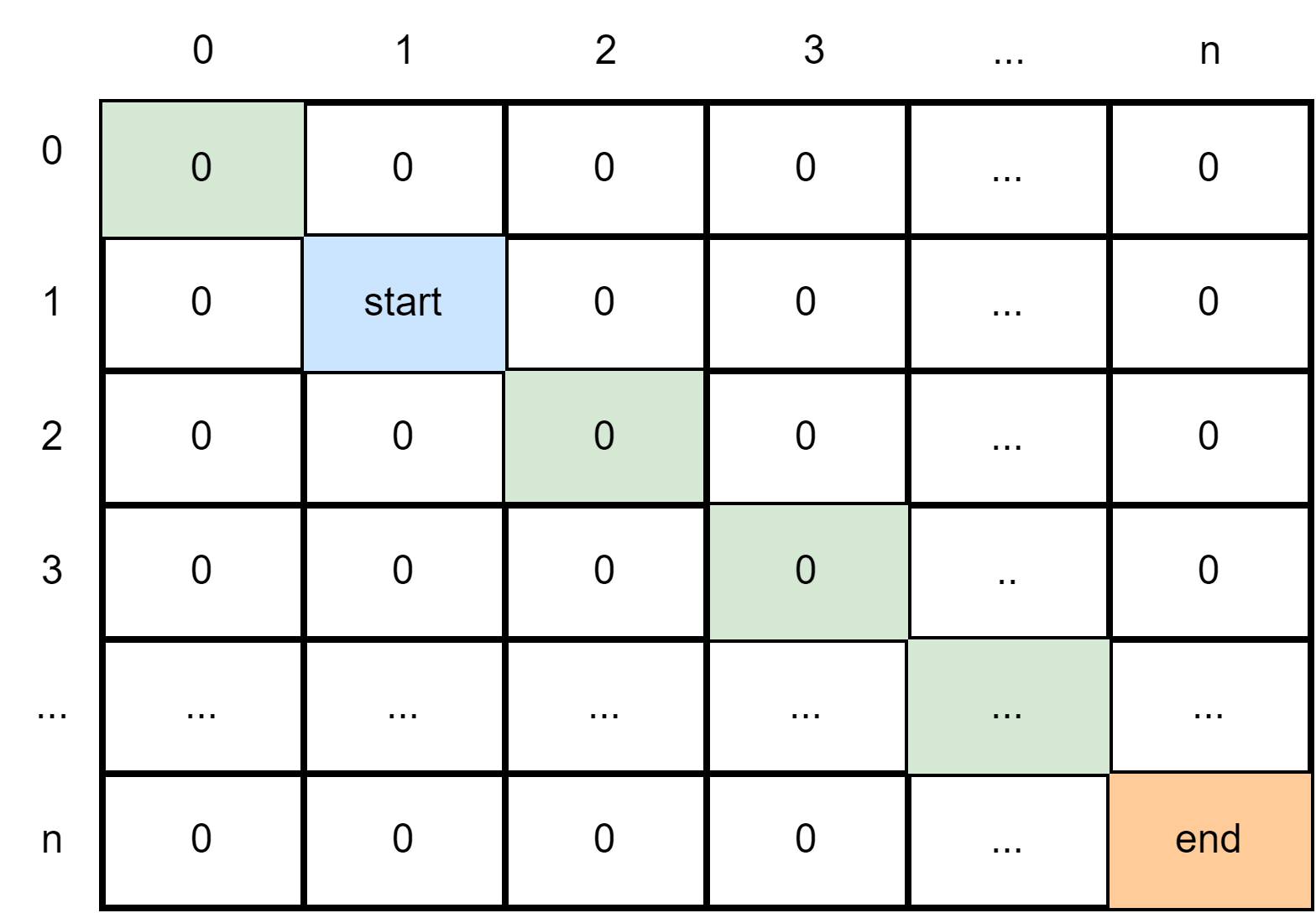}
	\caption{\textbf{ {$M_l$} position state matrix.}}
	\label{fig6}
\end{figure}

The state space describes the current state of the network environment as perceived by the agent, which includes the current location state $ S_l $ and the network state information $S_{info}$  of that location. The location state $S_l$ is composed of a two-dimensional matrix $M_l$ of size $H\times W$, as shown in Fig. \ref{fig6}. The main diagonal of this matrix represents the current effective location of the agent, which is specified as $M_{l}=\left [ \left ( x_{1} ,y_{1}  \right ),\left ( x_{2} ,y_{2}  \right ),...,\left ( x_{n} ,y_{n}  \right )  \right ]  $, where $n$ represents the number of network topology nodes. The starting point is represented by $ start $, the ending location is represented by $ end $, and the remaining elements of the matrix are represented by $ 0 $ as the location that the intelligent agent has not yet passed or will not pass. In the update process, the location information of the intelligent agent will be updated along with the currently selected action. When the next location state of the intelligent agent reaches the end position, the algorithm has completed the current iteration of routing decision-making.

\begin{figure}[t!]
	\centering
	\includegraphics[width=\linewidth]{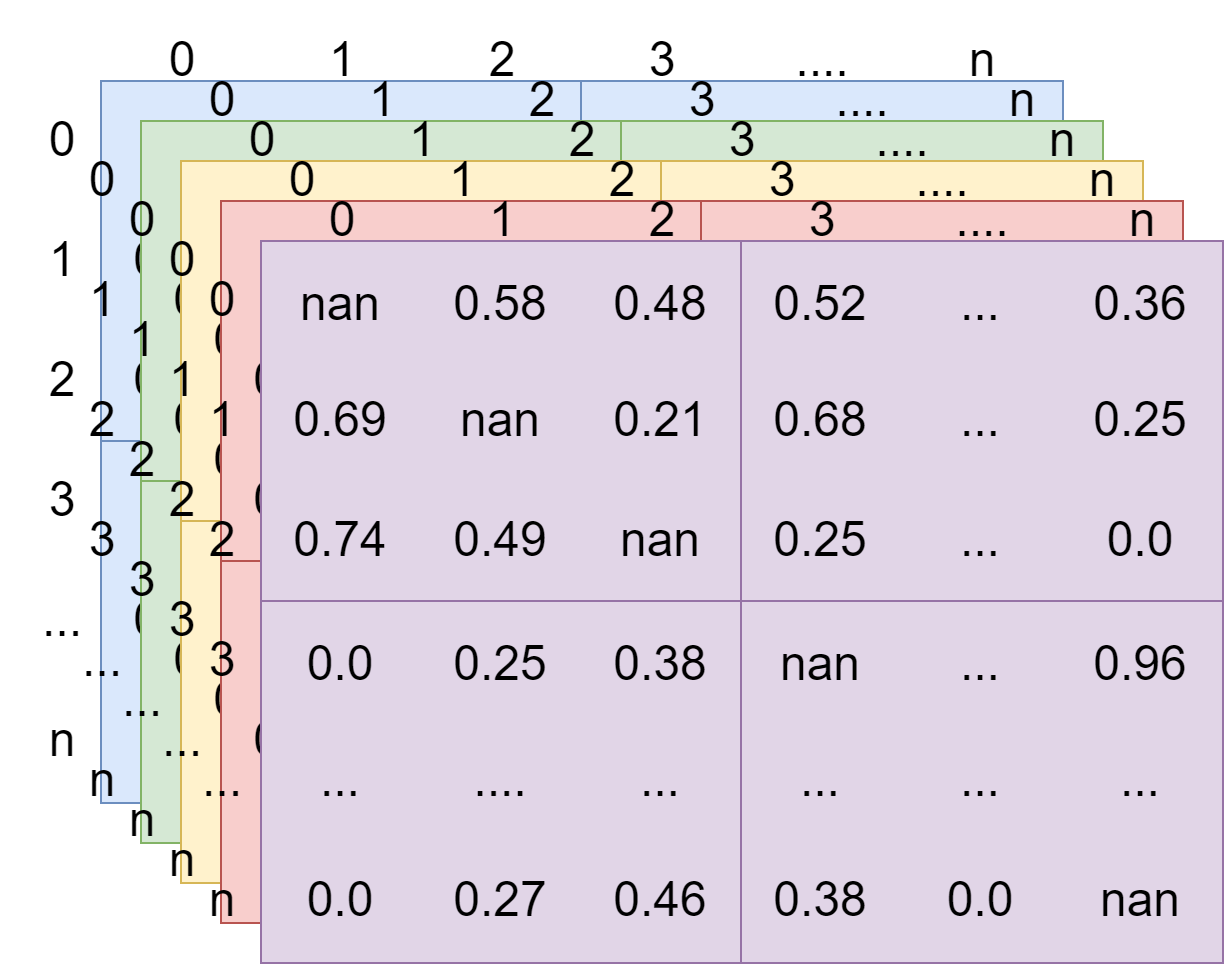}
	\caption{\textbf{{$M_{info}$} network information state matrix.}}
	\label{fig7}
\end{figure}

In this algorithm, the current network state information $S_{info}$ is composed of multiple two-dimensional state matrices $M_{ij}$  that contain various link information, such as the remaining bandwidth $bw_{free\left(ij\right)}$, the link delay $delay_{ij}$, the packet error rate $pkt_{err\left(ij\right)}$, the distance $distance_{ij}$ between wireless APs, and the packet loss rate $loss_{ij}$. The row number $ i $ and column number $ j $ of the matrix represent the identifying numbers of particular switch nodes; if $i\ne j$, the element in the $i$ row and $j$ column represents the link $e_{\left(i,j\right)}$ from node $i$ to node $j$; while if $i=j$, the element in the $i$ row and $j$ column represent node $i$ or $j$ itself. Here, $i,j\le n$, and $e_{\left(i,j\right)}=e_{\left(j,i\right)}\in E$, where $ E $ denotes the set of links in the network. The network state information matrix  $M_{info}$ is shown in Fig. \ref{fig7}. In this matrix, because the elements on the main diagonal do not correspond to network information between two switches, they are assigned an infinite $ nan $ value. Notably, the network link data obtained in an SWDN environment are unfavorable for training neural network models through gradient descent because the gradient explosion phenomenon tends to occur during the training process due to their large numerical values; this phenomenon reduces the model convergence speed of the algorithm and has a significant impact on the search trajectory of the agent. Therefore, the min-max method \cite{b43} is used to normalize the flow matrix, limiting the values of the elements in the matrix to within the range $\left [ a,b \right ] $, where $a,b \in \left[ {0,1} \right] $. In Equation \eqref{eq8}, $TM_{ij}$ represents the normalized flow matrix, $m_{ij}$ represents an element of the normalized flow matrix, as well as max(TM) and min(TM) represent the maximum and minimum values of the matrix elements, respectively. Due to the need for data preprocessing and the avoidance of calculation errors when the denominator is zero, a small numerical value is added to the denominator to ensure the accuracy of the calculation.

\begin{equation}
{\overline {TM} _{{\rm{ij}}}} = a + \frac{{\left( {{m_{ij}} - \min \left( {TM} \right) \cdot \left( {b - a} \right)} \right)}}{{\left( {\max \left( {TM} \right) - \min \left( {TM} \right) + 1{e^{ - 6}}} \right)}} \label{eq8}
\end{equation}

During the training process, to meet the input requirements of convolutional neural networks in reinforcement learning, the location state matrices $ M_l $ and the network state information matrices $M_{info}$ need to be concatenated in the channel dimension $C$ to form the input state space matrices. Namely, the two-dimensional matrix $M_l$ and $M_{info}$ of size $H\times W$ for $C$ channels are transformed into a three-dimensional spatial matrix tensor $T_l$ and $T_{info}$ of size $H\times W\times C$, where each channel corresponds to a two-dimensional matrix of $H\times W$.

\subsubsection{Action space $\mathcal{A}$}
	The action space describes the set of actions among which an intelligent agent must select during its interaction with the environment. After the intelligent agent executes an action based on the current state, the current state transitions to the next state. The design of the state space in this article is composed of location information and network information to enable the agent to quickly find the optimal path in the environment, improve the interaction between the agent and the environment, and reduce the dimensionality of the possible actions in the action space. In this article, each next-hop adjacent node is considered as a possible action node in the DRL action space, which is expressed as $A=\left \{ a_{1} ,a_{2},...,a_{i},a_{n-m},a_{n} \right \} \in \left \{ N_{1} ,N_{2},...,N_{i},N_{n-m},N_{n} \right \}$. Here, $ N_i $ represents a next-hop node, $ m $ represents the number of invalid actions, $ n $ represents the total number of wireless nodes, and $ a_i $ represents the current action being executed. In the process of action selection, there are three possible situations of the intelligent agent that need to be considered to provide corresponding decision-making strategies:
	\begin{itemize}
	\item If the next selected action node is not an adjacent node of the current node, the selected node will not be added to the selected path matrix. The agent will remain in its current location, and the behavior of the agent will be punished;
	
	\item If the selected action node belongs to the set of adjacent nodes but will cause the paths in the path matrix to form a loop, the agent will also be punished to some extent. Nevertheless, the location state of the intelligent agent will be changed, and the link information matrix $T_{info}$ will be updated; 
	
	\item When the agent selects the endpoint node, its location state will change, and a reward value should be returned to incentivize the intelligent agent to select this path, thus completing the current iteration of the routing decision-making process.
	\end{itemize}
\subsubsection{Reward function $\mathcal{R}$}
	The reward function is used to optimize the direction of agent learning. In the algorithm designed in this article, the agent always makes routing decisions toward the direction of the maximum reward value, comprehensively considering the overall performance of the network to achieve multi-objective routing optimization. Therefore, the reward value is calculated using the remaining link bandwidth $bw_{free(ij}) $, the link delay $delay_{ij} $, the link packet error rate $pkt_{err(ij} $, the distance $distance_{ij} $ between wireless APs, and the link packet loss rate $loss_{ij} $ in the SDWN network. Accordingly, the reward value between two nodes, also called the reward value for the next hop, can be represented by a two-dimensional matrix $R_{link} $, and the final reward value for the entire link can be represented by a two-dimensional matrix $R_{total} $. The former allows the agent to make local routing decisions, while the latter allows the agent to make global routing decisions. The purpose of designing these two types of reward functions is to prevent the agent from falling into locally optimal solutions and enable the agent to quickly find the optimal path. The relationship between the two rewards is represented by Equation \ref{eq9}, where the row number $i$ and column number $j$ of the matrix represent the identifiers of particular switch nodes.

\begin{equation}
{R_{total}} = \sum\limits_{i,j = 0}^n {{R_{link\left( {ij} \right)}}}\label{eq9}
\end{equation}

During its interaction with the environment, an intelligent agent may encounter the following three situations, and the reward value for the next hop will be split into three different rewards accordingly. The intelligent agent will receive the corresponding reward value based on the current environmental situation for cumulative learning.

	\begin{itemize}
	\item 	When the agent selects an adjacent node of the current node as the next action, the path it passes through does not form a loop, and the next state is not the destination state, that is, $S_{t+1}\ne None $, the reward value obtained is $R_{link}$. This reward value is calculated as shown in Equation \eqref{eq10}, where  $\beta_{1},\beta_{2} ,\beta_{3} ,\beta_{4} ,\beta_{5}   \in \left[0,1\right]$. 	The link performance indicators that will be prioritized for optimization can be controlled by adjusting the weight values $\beta$.
	
	\begin{equation}
	\begin{array}{c}
	{R_{link}} = {\beta _1}{\overline {bw} _{ij}} - {\beta _2}{\overline {delay} _{ij}} - {\beta _3}{\overline {pkt} _{err\left( {ij} \right)}} \\
	\\
	-{\beta _4}{\overline {distance} _{ij}} - {\beta _5}{\overline {loss} _{ij}}
	\end{array} \label{eq10}
	\end{equation}
	
	\item 	When the next action selected by the agent is an adjacent node of the current node and forms a loop in the path traveled by the agent, the reward value obtained is $R_{loop}$. This reward value is given in the Equation \eqref{eq11}, where $\xi _{1} $ is the discount factor , the negative sign indicates that the agent is penalized with a negative value to discourage it from performing this action again, and $R_s $ is the standard reward; $\xi _{1} ,R_s \in \left[0,1\right]$. By varying the size of the discount factor to optimize the decision-making direction of the agent, the agent's motion trajectory can be made more stable, and the agent's routing policy can be dynamically adjusted according to different network topologies.
	
	\begin{equation}
     {R_{loop}} =  - {\xi _1}{R_s} \label{eq11}
	\end{equation}
	
	\item 	When the agent selects an action node that is not adjacent to the current node, the reward value obtained is $R_{non\_edges}$. As shown in Equation \eqref{eq12}, the standard reward $R_s $ is multiplied by a discount factor $\xi_2 $, where $\xi _{2} ,R_s \in \left[0,1\right]$, and the negative value is taken to penalize the selection of invalid actions and increase the probability of the agent selecting correct actions.
	 
	\begin{equation}
	{R_{non\_edge{\rm{s}}}} =  - {\xi _2}{R_{\rm{s}}} \label{eq12}
	\end{equation}
	
	\end{itemize}
\subsubsection{Importance sampling}
The general gradient descent update algorithm is used because the value estimated by the advantage evaluation function is not completely accurate, so there will be deviations in the data. If the data obtained by the agent deviates too far from the actual value after the execution of a policy, the next sample will also strongly deviate from the estimate, resulting in the agent executing a policy that also strongly deviates from the predicted policy. At the same time, because of the different distribution of the data used for training, the network parameters $\theta$ will change to $\theta ^{'} $ after learning. Therefore, the importance sampling method is used in this paper to adjust the distribution of the data, avoid problems caused by data estimate deviation, and improve the efficiency of policy updates and sample utilization. Specifically, Equation \ref{eq13} is importance sampling, where $w_{t}^{IS} $ is the weight of importance sampling, and $\pi _\theta \left ( {a_t}|{s_t}  \right ) $ and $\pi _{\theta old} \left ( {a_t}|{s_t}  \right ) $ represent the probabilities of taking action $a_t $ in state $s_t $ when using the new policy and the old policy, respectively.
	\begin{equation}
	w_t^{IS} = \frac{{{\pi _\theta }\left( {{a_t}|{s_t}} \right)}}{{{\pi _{{\theta _{old}}}}\left( {{a_t}|{s_t}} \right)}} \label{eq13}
	\end{equation}
	
\subsubsection{Gradient clipping optimization}
 During importance sampling, to avoid a large difference between the original data distribution $p(x)$ and the data distribution $q(x)$ that can be sampled, it is necessary to add a constraint $\sigma$ on the distance between the two, as represented by the KL divergence. Because the loss function of the PPO reinforcement learning algorithm is constrained by relevant conditions and calculated via the conjugate gradient method and because the KL divergence constraint, as an additional constraint condition, does not participate in the update of the convolutional neural network parameters, it is difficult to dynamically adjust the parameters in the KL divergence to adapt to different data distributions. Therefore, the KL divergence is introduced into the loss function as a regularization term for optimization. The gradient descent formula with the KL divergence regularization term is shown in the Equation \eqref{eq14}. 

\begin{equation}
\begin{gathered}
J_{PPO}^{{\theta ^k}} = {J^{{\theta ^k}}}\left( \theta  \right) - \sigma KL\left( {\theta ,{\theta ^k}} \right) \hfill\\
{J^{{\theta ^k}}}\left( \theta  \right) \approx  \sum\nolimits_{\left( {{s_t},{a_t}} \right)} {\frac{{{p_\theta }\left( {{a_t}|{{\text{s}}_t}} \right)}}{{{p_{{\theta ^k}}}\left( {{a_t}|{{\text{s}}_t}} \right)}}} {A^{{\theta ^k}}}\left( {{s_t},{a_t}} \right) \hfill\\
{A^{{\theta ^k}}}\left( {{s_t},{a_t}} \right) = {R_t} + \gamma V\left( {{s_{t + 1}}} \right) - V\left( {{s_t}} \right) \hfill \\ 
\end{gathered}
 \label{eq14}
\end{equation}
where $\theta$ represents the policy parameters in the neural network; $\sigma$ represents the weight coefficient of the KL divergence, satisfying $\sigma \in \left [ 0,1 \right ] $;  $\theta^k$ represents the network parameters after $k$ iterations; $J^{{\theta ^k}}$ represents the likelihood function of gradient decline after $k$ iterations; ${{{p_{{\theta }}}\left( {{a_t}|{{\rm{s}}_t}} \right)}}$ represents the action–state transition probability distribution function; ${A^{{\theta ^k}}}\left( {{s_t},{a_t}} \right)$  represents the advantage evaluation function after $k$ iterations, which is used to evaluate the value of the current strategy of the agent; $R_t $ refers to the reward function at time $t$; $\gamma$ is a reward attenuation factor; and $V\left( {{s_{t}}} \right)$ and $V\left( {{s_{t + 1}}} \right)$ are the estimated values for the current state and the next state, respectively.

Due to the computational complexity of the KL divergence \cite{b44} formula and the difficulty of selecting an appropriate penalty factor $ \sigma $ to adjust the similarity between the old and new policies, to better adapt to dynamic network topology changes and uneven data distributions while effectively solving the problem of gradient explosion or vanishing, this algorithm uses the gradient clipping method in place of the KL divergence. Gradient clipping not only effectively limits the policy update amplitude but also improves the model convergence speed and performance of the algorithm. The formula for the gradient clipping method is given in \eqref{eq15}.

\begin{equation}
\begin{gathered}
cli{p_t}\left( \theta  \right) = \min \left( {\max \left( {{\eta _t}\left( \theta  \right),1 - \epsilon ,1 + \epsilon } \right)} \right) \\ 
{\eta _t}\left( \theta  \right) = \frac{{{\pi _\theta }\left( {{a_t}|{s_t}} \right)}}{{{\pi _{{\theta _{old}}}}\left( {{a_t}|{s_t}} \right)}} \\ 
\end{gathered}  \label{eq15}
\end{equation}
where $\theta$  represents the policy parameters in the neural network,  $ \epsilon \in \left[0,1\right]$ is the clipping coefficient used to adjust the amplitude of the policy update, and $ \eta _t (\theta) $ is the ratio of the new policy $ \pi_\theta (a_t |s_t ) $  to the old policy $  \pi_{\theta old} (a_t |s_t ) $ at time $t$.

Gradient clipping limits the ratio between the new policy function $ \pi_\theta (a_t |s_t ) $ and the old policy function $  \pi_{\theta old} (a_t |s_t ) $  to a range of $ (1-\epsilon,1+\epsilon) $. If the gradient norm exceeds this range, clipping is applied, and finally, the minimum value is taken as the output result.

\subsubsection{PPO update strategy}	
The policy update formula of this algorithm maximizes the upper limit on the expected cumulative return while limiting the amplitude of the policy updates to ensure the convergence and stability of the algorithm. The policy update formula of the PPO algorithm is shown in Equation \eqref{eq16}.

\begin{equation}
\begin{gathered}
{\theta _{t + 1}} =  \hfill \\
argma{x_\theta }\widehat {{E_t}}\left[ {\min \left( \begin{gathered}
	{\eta _t}\left( \theta  \right){A^{{\theta ^k}}}\left( {{s_t},{a_t}} \right), \hfill \\
	clip\left( \begin{gathered}
	{\eta _t}\left( \theta  \right), \hfill \\
	1 - \epsilon , \hfill \\
	1 + \epsilon  \hfill \\ 
	\end{gathered}  \right){A^{{\theta ^k}}}\left( {{s_t},{a_t}} \right) \hfill \\ 
	\end{gathered}  \right)} \right] \hfill \\ 
\end{gathered}\label{eq16}
\end{equation}
where $ \widehat {{E_t}} $ represents the expected value at time $ t $, $ {\eta _t}\left( \theta  \right) $ is the ratio of the new and old policies at time $ t $, and $  A^{\theta^{k}  }  (s_t,a_t ) $ is the advantage evaluation function for executing action $ a_t $ after $ k $ iterations, used to measure the quality of the executed policy. The function $ clip(r_t (\theta),1-\epsilon,1+\epsilon) $ ensures that the update does not deviate too far from the old policy.

\subsection{Description of the DRL-PPONSA Algorithm}
The implementation of the DRL-PPONSA algorithm framework is shown in Algorithm \ref{alg:algorithm2}. Based on the input source node set $N_s $ and destination node set $N_d $, the optimal path from $N_s $ to $N_d $ is found from the currently observed network environment topology $G$. The prediction matrix output by Algorithm \ref{alg:algorithm1} is used as the network state information matrix that is also provided as input to Algorithm \ref{alg:algorithm2}. In addition, the input to the algorithm includes the location state matrix $M_l $ corresponding to the current network state, the learning rate $\alpha $, the network parameter update frequency $ f_{update} $, the size $ \mathcal{K} $ of each batch drawn from the experience pool, the hyperparameter $\epsilon $ used in importance sampling, and the number of training episodes $\mathcal{M}$.

Lines 1–4 mainly initialize the parameters of the policy function $\pi_{\theta} (s)$ and the value function $V_\psi (s)$, while leaving the experience pool empty. In lines 7 and 8, the location state matrix $M_l $ and network state information matrix $M_{info} $ are reset based on a randomly selected source node $N_s $ and destination node $N_d $. Then, the two matrices are concatenated in the channel dimension to form matrices $T_l $ and $T_{info} $ with dimensions of $N\times N\times C\times \mathcal{K}$ as the initial state $S_t $, where $n$ is the number of nodes, that is, the size of the action space, and $\mathcal{K}$ is the batch size. Then, training is started in environments with different remaining bandwidths, delays, packet loss rates, packet error rates, and link transmission distances. Lines 5–35 describe one iteration of the training process, in which the agent travels from the source node to the destination node and outputs the selected path.

From line 9 to line 14, the agent executes the policy function $\pi_\theta (s)$ to obtain the current state matrix $s_t $, the selected action set $a_t $, and the probability $p(a_t )$ of selecting each action as well as the total cumulative reward value $r_t $ in the current environment. The advantage function $ \widehat {{E_t}} $ is then evaluated, and based on the reward, the action policy is updated. Finally, the obtained next state value, current state value, and cumulative reward are stored in the experience pool to be used to train the parameters of the neural network.

Lines 15 to 22 describe the process of updating the parameters of the algorithm, mainly consisting of the loss function computation and update processes for the critic and actor networks. First, relevant data are retrieved from the experience pool, including the current state matrix $s_t $, the selected action and its corresponding probability $p(a_t )$ from the policy distribution, the cumulative reward value $r_t $ of the agent, and information regarding the state matrix $s_{t+1}$ that the agent updates. The previously initialized advantage function $ \widehat {{E_t}} $ is updated, and then the loss function of the policy network is updated based on the next-action probability $p(a_{t+1} )$ and the previous-action probability $p(a_t )$, using the likelihood function $J_{PPO}^{{\theta ^k}}(\theta)$ to update the network weights. Finally, the cumulative reward values are used to calculate the loss value for evaluating the network, guiding the agent to prioritize the path with the maximum cumulative reward.

At the end of lines 23 to 38, it is judged whether the agent has reached the destination. If the next state is not the destination, the cumulative reward value is calculated. The calculated reward value is the discounted link reward. If the agent has reached the destination, the target reward value is received. At the same time, the parameters of the two networks are updated in accordance with the likelihood function. When the agent reaches the destination, the current iteration cycle ends, and the status of the intelligent agent is updated to output the currently selected path. After the path is updated, a reward $r_t $ and new network state $s_{t+1} $ will be obtained. In this way, the performance of the network is optimized toward the maximum reward value until the optimal routing policy is found.

\begin{algorithm}[t!]
	\small
	\caption{DRL-PPONSA}
	\label{alg:algorithm2}
	\begin{algorithmic}[1]
		\Require location information matrix $M_l$,
		network status information matrix $M_{info}$,
		source nodes set  $N_s$,
		destination nodes set $N_d$,
		learning rate $\alpha  $,
		n-step $n_{step}$,
		batch-size $\mathcal{K}$,
		clip-param  $\epsilon$,
		ppo-update-time $n_{update}$,
		training episodes $\mathcal{M}$.
		\Ensure
		Intelligent path information for $(N_s, N_d)$.
		\State \textbf{Initialize} actor network with random weight $\theta$.
		\State \textbf{Initialize} critic network with random weight $\psi $.
		\State \textbf{Initialize} PPO policy function $\pi_\theta \left(s\right)$ and value function $V_\psi \left(s\right) $.
		\State \textbf{Initialize} n-steps buffer-capacity $B$.
		\For{$episode=1\leftarrow \mathcal{M}$}
			\For{$M_{bw_{free}},M_{delay},M_{loss},M_{pkt_{error}},M_{distance}$ 
				 in Network \hspace*{2.0em}Information Storage }
			\State Reset enviroment with $(N_s, N_d)$.
			\State Get $s_t \leftarrow stack(M_l,M_{info})\leftarrow(T_l,T_{info})$.
			 //concatenate \hspace*{3.0em} in the channel dimension
			 \While{True}
			\State Run policy $\pi _{\theta } $ for  $n_{update}$, collecting {$\left(s_{t},a_{t},r_{t},p\left(a_{t}\right)\right)$}
			\State Estimate advantages $\widehat{A_t} =\sum _{t^\prime >t}\gamma^{t^\prime -t} r_{t^\prime} -V_{\phi } \left(s_{t}  \right)$;
			\State Update the strategy of the action $\pi _{\text{old}} \longleftarrow \pi _{\theta}$; 
			\State Execute action $a_t$ and observe reward $r_t$ and next \hspace*{4.5em} state $s_{t+1};$
			\State Store $(s_t, a_t, r_{t},p\left(a_{t}\right), s_{t+1})$ in $B.$
			
			\For{$i=1\leftarrow n_{update}$}
			\For{index in $\left(range\left(len(\mathcal{K}\right)),\mathcal{K},False\right)$}
			\State Sample minibatch {$(s_i, a_i, r_{i:i+n}, s_{i+n})$} and get \hspace*{7.5em} $p\left(a_{t}\right)$ from Transition
			\State Update the advantages $\widehat{A_t}$ and calculate the \hspace*{7.5em} qualifying ratio $r=\frac{p\left(a_{t}\right)}{p\left(a_{t+1}\right)}$
			\State Calculate the likelihood function of gradient \hspace*{6.5em} descent 
			$J_{\text{PPO-Clip}}^{\theta^{k}}$ //The Equation is given by \eqref{eq14}
			\State Update $\theta$ by a gradient method w.r.t $J_{\text{PPO-Clip}}^{\theta^{k}}$
			\State Calculate the actor loss and critic loss.
			\EndFor
			
			\If{$s_{i+n}$ is not None}
			\State $R\leftarrow R_{i:i+n} + \gamma\sum\limits_{t=1}^{T}R_{link(t)}$
			\Else
			\State $R\leftarrow R_{i:i+n}$
			\EndIf
			
			\State Update policy network parameters  $\theta \gets \theta +\hspace*{6.5em} \zeta_{\theta }  \nabla J_{\text{PPO-Clip}}^{\theta^{k}}\left(\theta\right)$ 
			\State Update critic network parameters $\psi  \gets \psi + \hspace*{6.5em} \zeta_{\psi }   \nabla J_{\text{PPO}}^{\psi^{k}}\left(\psi\right)$
			\EndFor
			
			\If{The agent arrives at the destination node}
			\State Break
			\EndIf 
			
			\State $s_t \leftarrow s_{t+1}$
			
			\EndWhile
			\State Output the path selected by the agent from $\left(N_s, N_d\right)$
		\EndFor
	\EndFor	
	\end{algorithmic}
\end{algorithm}

\section{EXPERIMENTAL SETTINGS AND PERFORMANCE EVALUATION}
This section mainly introduces the parameter settings and flow generation methods of the simulation environment. Then, the performance of the prediction algorithm is analyzed, and a parameter comparison analysis of the DRL algorithm is presented. Finally, the results of the proposed algorithm are compared with those of Dueling DQN, OSPF, DVRP, and LSRP.

\subsection{Simulation Environment Settings}
The control plane used in these experiments uses Ryu as the SDWN controller, which is responsible for matching and issuing flow table entries and responding to events. For the data plane, the Mininet-WIFI 2.3.1b \cite{b50} simulation environment is used to build a simulated wireless network topology, and the sFlow-RT tool \cite{b45} is used to monitor the flow situation in the network. The entire simulation environment is deployed on an Ubuntu 18.04.6 server equipped with a GeForce RTX 3090 graphics card. Ryu \cite{b48} uses the southbound interface of the OpenFlow 1.3 protocol to communicate with Mininet-WIFI's Open vSwitches \cite{b46}. The dataset is collected by Ryu and stored as a pickle file in graph format. Finally, the interaction between SDWN and reinforcement learning is achieved using Python 3.8 and Pytorch 1.11.0. The simulation tools are listed in Table \ref{tab1}.

\begin{table}[htb] 
	\renewcommand\arraystretch{1.5}
	\centering  
	\caption{\textbf{Simulation tools}}
	\label{table}
	\setlength{\tabcolsep}{6pt}
     \begin{tabularx}{\linewidth}{|>{\centering\arraybackslash}X|>{\centering\arraybackslash}X|>{\centering\arraybackslash}X|}
		\hline
		Tools&
		Version&
	    Function \\
		\hline
		Mininet-WIFI& 2.3.1b& network topology construction \\
	    
	    Ryu& 4.3.4& flow table distribution\\
	    
	    GPU& GeForce RTX3090& accelerated calculation\\
	    
	    Ubuntu&	18.04.6& experimental system\\
	    
	    sFlow-RT&	3.0&	flow monitoring\\

		\hline
	\end{tabularx}
	\label{tab1}
\end{table}

\begin{figure}[t!]
	\centering
	\includegraphics[width=\linewidth]{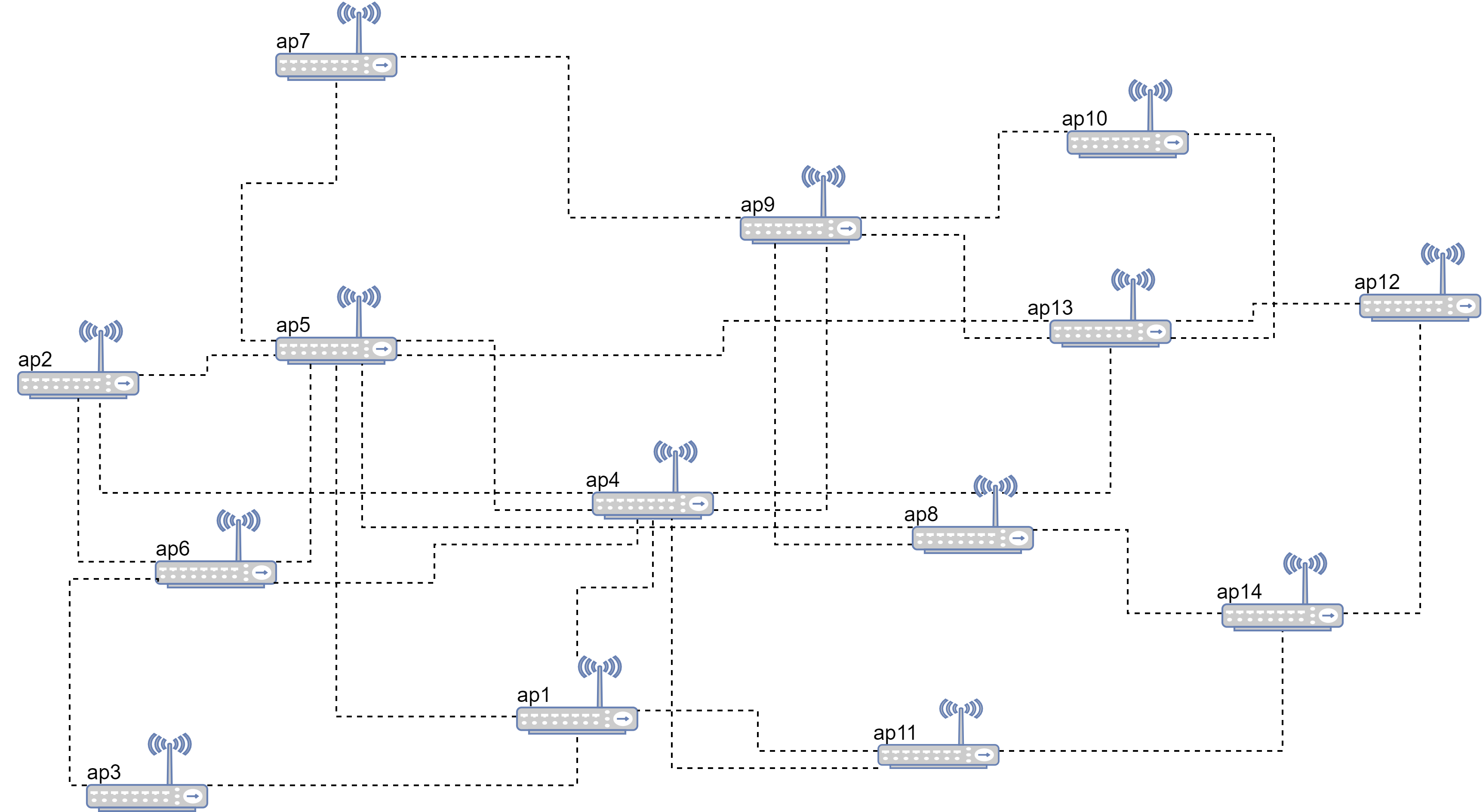}
	\caption{\textbf{The wireless network topology of APs .}}
	\label{fig8}
\end{figure}

The experimental topology built using the Mininet-WIFI simulation platform is shown in Fig. \ref{fig8}. It is a simulated wireless IoT data center topology that includes $ 14 $ nodes and $ 25 $ links. The dashed lines in this figure represent wireless links between wireless APs, and each wireless AP is connected to a STA. Table \ref{tab2} provides information such as the simulation range, simulation time, and basic parameters of the simulation environment between the wireless APs and STAs. The link parameters between wireless APs are randomly generated following uniform distributions. The bandwidth is set  to $ 5\sim 40 Mbps $, the delay is $  1\sim10 ms $, the link packet loss rate is set to $ 0.1\sim 1 \% $, and the distance between wireless APs is set to $ 30\sim 110 $ meters.

\begin{table}[htbp] 
	\renewcommand\arraystretch{1.5}
	\centering  
	\caption{\textbf{Simulation parameters}}
	\label{table}
	\setlength{\tabcolsep}{6pt}
	\begin{tabularx}{\linewidth}{|>{\centering\arraybackslash}X|>{\centering\arraybackslash}X|>{\centering\arraybackslash}X|>{\centering\arraybackslash}X|}
		\hline
		Parameters&	Value&	Parameters&	Value \\
		\hline
		MAC protocol&	IEEE 802.11g&	Frequency band&	$ 2.4 GHz $\\
		Number of APs&	$ 14 $(nodes)&	Number of STAs&	$ 14 $(nodes)\\
		CCA threshold&	$ -62 dBm $&	Channel&	$ 1,6  $\\
		Propagation loss& 	log-distance&	Transmission power&	$ 21 dBm $ \\
		Receiving gain&	$ 5 dBm $&	Transmitting gain&	$ 5 dBm $ \\
		Signal range&	$ 38\sim140 m $&	Simulation area&$ 	300x300 m\textsuperscript{2} $ \\
		Modulation technique&	OFDM&	Simulation time&	$ 180 min $ \\
		
		\hline
	\end{tabularx}
	\label{tab2}
\end{table}

\begin{figure*}[htp]
	\centering
	\includegraphics[width=\linewidth]{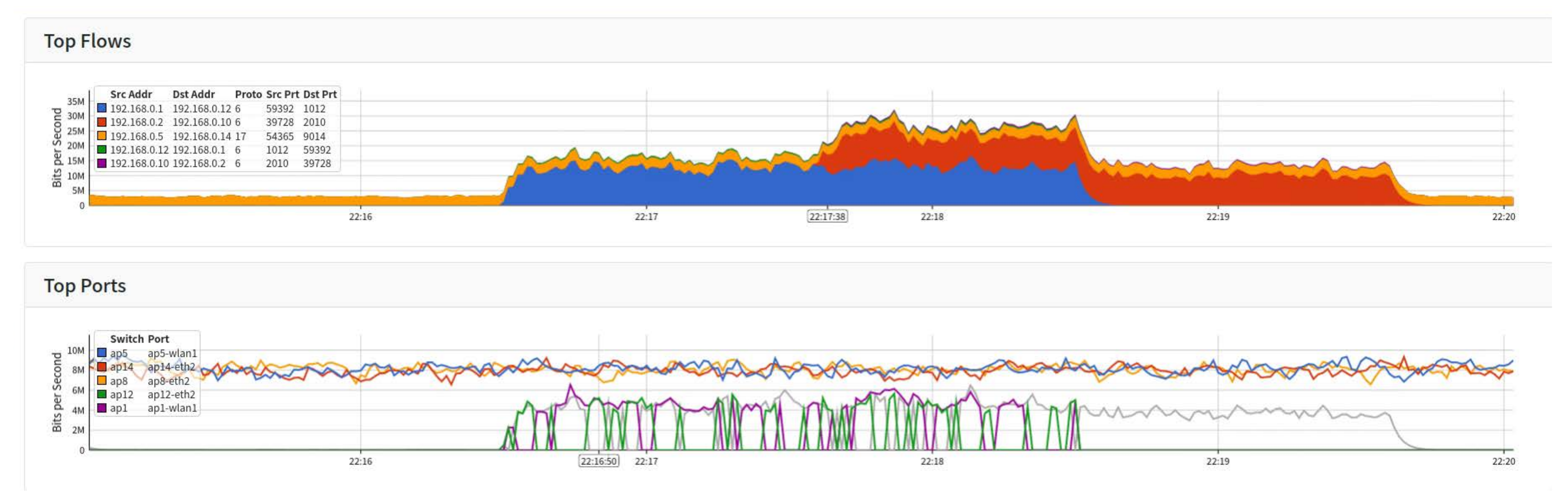}
	\caption{\textbf{The curve of port traffic information varies with time.}\label{fig9}}
\end{figure*}

To simulate real wireless network traffic usage, we use the iPerf3 \cite{b49} tool to write Python scripts that generate traffic information. A script sends User Datagram Protocol (UDP) traffic requests to a server through a client, randomly selecting the client and server for multi-objective traffic and adjusting the size of the traffic based on the gravity model \cite{b48}. The flow size is controlled within the range of $0\sim50 MHz$  to ensure that the flow information reaches its peak during the 10:00-15:00  period of the day and slowly decreases during other periods. The traffic information collection interval is once every $ 5 $ s. Finally, the Ryu controller monitoring module script generates 1000 instances of graph data for $ 25 $ traffic matrices between the $ 14 $ nodes and writes the graph data to a pickle file. The elements in the matrices include link information such as residual bandwidth, used bandwidth, delay, packet loss rate, and distance. The sFlow-RT tool \cite{b51} is used to monitor the average traffic flow information in bits per second, as shown in Fig. \ref{fig9}.

\subsection{Analysis of GCN-GRU Prediction Network Performance and DRL Parameters}
\begin{figure*}[htp]
	\centering
	\begin{minipage}[t]{0.48\linewidth}
	\subfigure[The curve of reward varies with episode]{
		\centering
		\includegraphics[height=5.54cm,width=\linewidth]{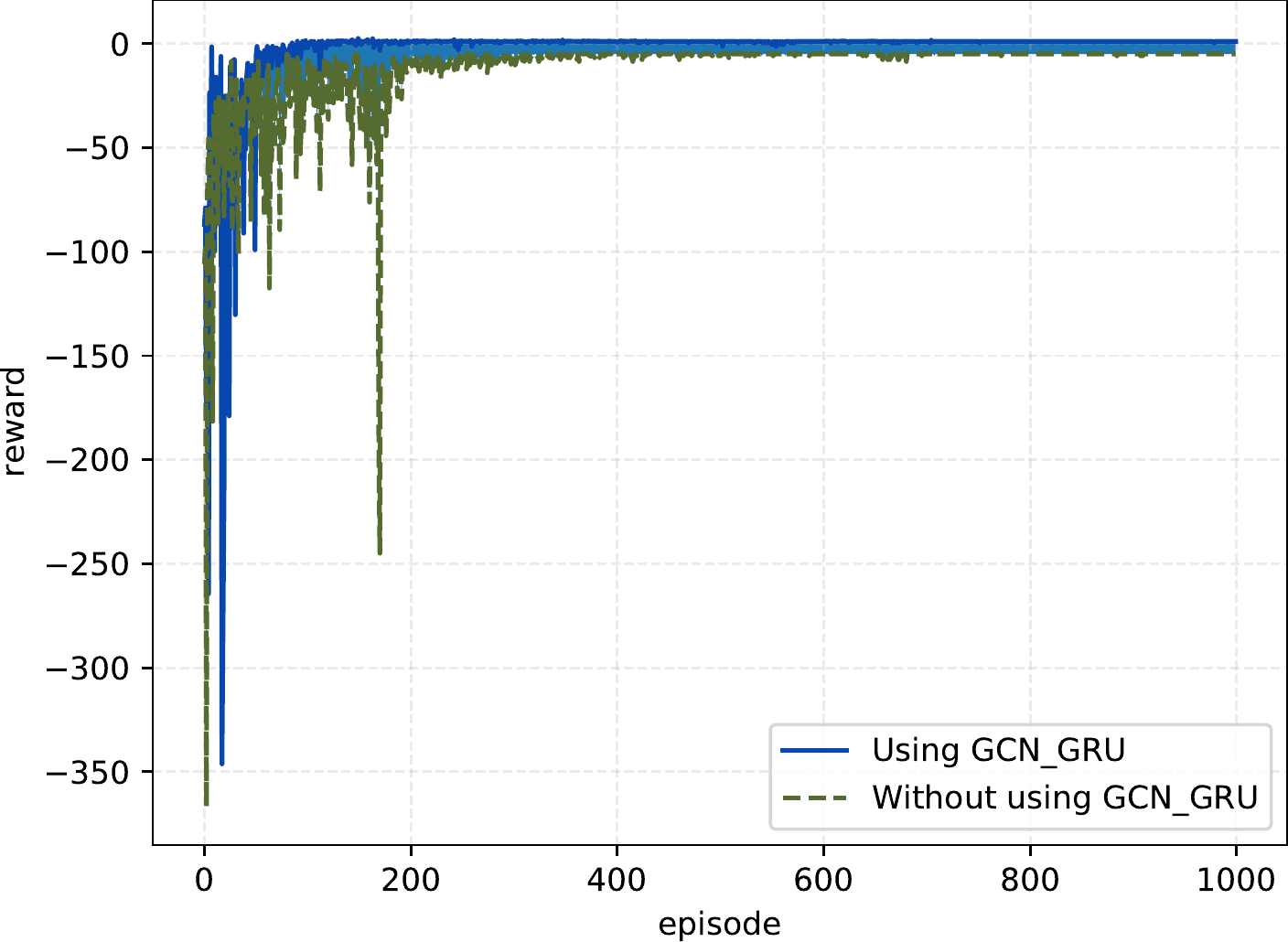}
     	\label{subfig:10(a)}
           }
	\end{minipage}
	\hfill
	\begin{minipage}[t]{0.48\linewidth}
	\subfigure[The curve of step varies with episode]{\centering
		\includegraphics[height=5.54cm,width=\linewidth]{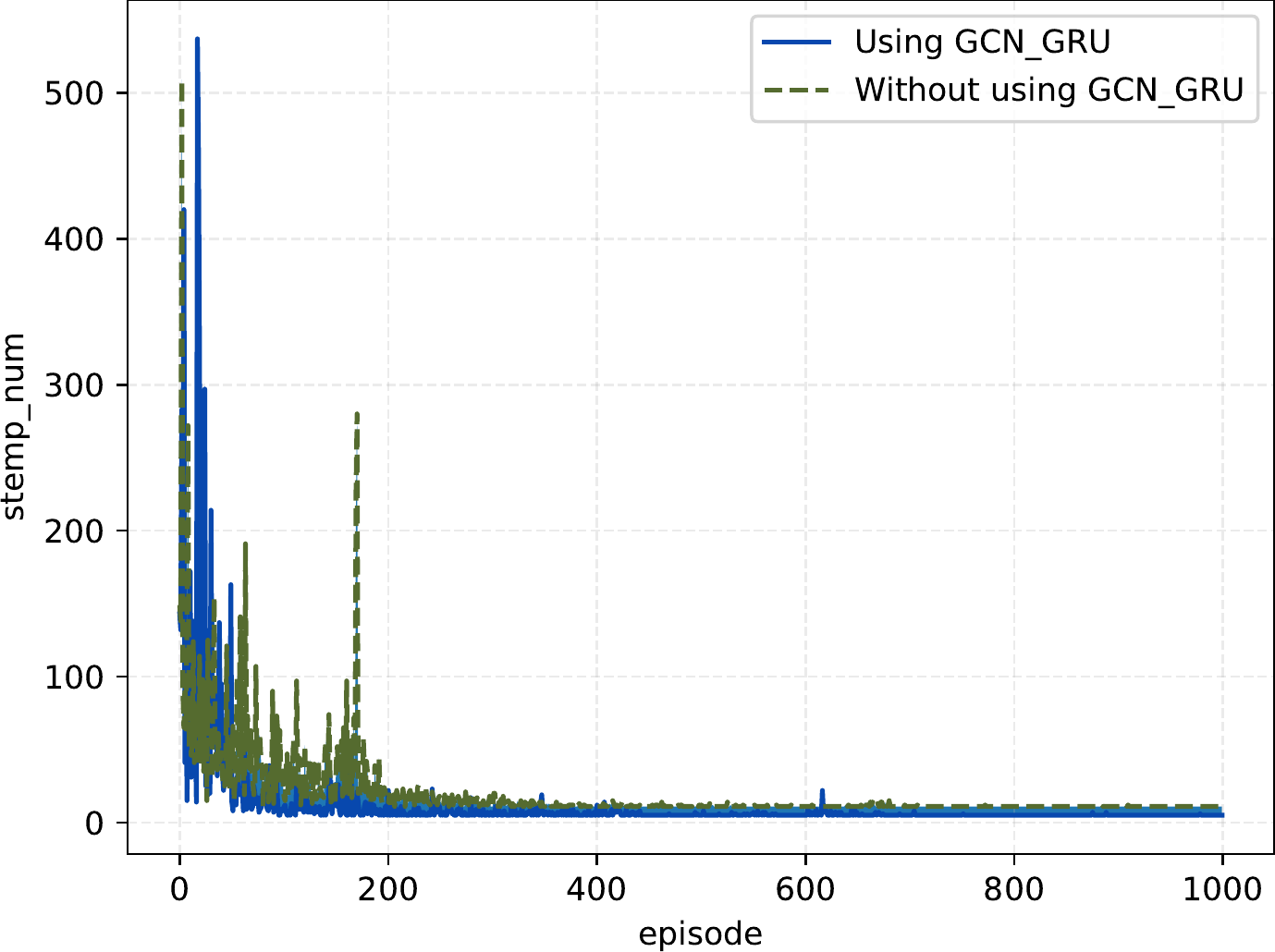}
		\label{subfig:10(b)}
            	}
	\end{minipage}
     \caption{\textbf{Comparing reward and step change curves with and without the Implementation of GCN-GRU prediction mechanism.}\label{fig10}}
\end{figure*}

The GCN-GRU model fully utilizes the capabilities of GCNs to obtain spatial graph information and utilizes the advantages of GRU networks for time series modeling to mine hidden and unknown traffic information from SDWN networks, thereby improving the perception ability of the DRL algorithm and making the generated routing policies more forward-looking and robust. As shown in Fig. \ref{subfig:10(a)}, the reward values obtained by an agent using the GCN-GRU prediction mechanism are significantly higher than those obtained without the prediction mechanism. While in Fig. \ref{subfig:10(b)}, the number of steps taken by the agent using the prediction mechanism is also significantly fewer than that taken by the agent without the prediction mechanism. This is because the prediction mechanism can capture unknown traffic status information in the network, allowing the intelligent agent to perceive more traffic information, make judgments in advance for routing decisions, and avoid selecting congested nodes. Moreover, when relying solely on Ryu's network measurement mechanism, it is difficult to detect hidden traffic information in the SDWN network, and it is also difficult to adjust policies promptly to adapt to the distribution of traffic. Therefore, the proposed traffic prediction algorithm based on the GCN-GRU model can enhance the exploration space of the DRL-PPO intelligent routing algorithm to optimize the action selection and increase the reward value, thus yielding more appropriate routing policies for SDWN network flows and improving the utilization of network resources. These findings verify that the proposed prediction mechanism can improve the performance of intelligent routing algorithms.

The intelligent routing algorithm designed in this article adopts the PPO algorithm, with updates based on policy gradient descent. It uses an online learning strategy, in which the same agent is used in both the training stage and the policy execution stage. During the learning process, the intelligent agent can complete policy selection for the next state, and the selection for the previous state does not affect the policy selection for the next state, while the opposite is true in offline learning. The basic parameters set for the intelligent agent during the training process are shown in Table \ref{tab3}. For PPO reinforcement learning, a four-layer network structure is adopted, which includes a two-layer actor network and a two-layer critic network. The actor network is used to sample the probability of the current action and the probability of the next state's action, while the critic network is used to evaluate the cumulative reward of the current state and the cumulative reward of the next state. For the optimizer, Adaptive moment estimation (Adam) is used, which can adaptively update its learning rate and reduce the consumption of computer memory, enabling the network model of the intelligent routing algorithm to quickly converge during the training process.

\begin{table}[htbp] 
	\renewcommand\arraystretch{1.5}
	\centering  
	\caption{\textbf{DRL parameter settings}}
	\label{table}
	\setlength{\tabcolsep}{6pt}
	\begin{tabularx}{\linewidth}{|>{\centering\arraybackslash}X|>{\centering\arraybackslash}X|>{\centering\arraybackslash}X|}
		\hline
     	Parameter&	Symbol&	Value \\
		\hline
		Learning rates&	$ \alpha_{1},\alpha_{2} $&	$ 1e^{-3} ,3e^{-3}  $ \\
		Bach size &       $ k $ &  $ 32 $ \\
	Clip parameter&	      $ \epsilon $& 	$ 0.2 $\\
	PPO-update-time&	   $ f $&		$ 10 $ \\
	Buffer-capacity& 	$ B $&	     $ 3000 $ \\
		Reward attenuation&	$ \gamma $&		$ 0.99 $ \\
		Discount factors&  $ \xi_{1},\xi_{2} $&  $ 0.5,0.8 $\\
		episodes&  $ M $ &          $  1000 $ \\

		\hline
	\end{tabularx}
	\label{tab3}
\end{table}

Choosing appropriate weight factors will improve the model convergence effect of the intelligent routing algorithm and accelerate the convergence process. The routing algorithm in this article is an intelligent routing algorithm aimed at optimizing the parameters of multiple target links, which includes maximizing the remaining bandwidth and minimizing the delay, packet loss rate, packet error rate, and transmission distance. In the experimental process, we choose the default weights $[0.7,0.3,0.1,0.1,0.1]$ for the optimization objective, which means that $ \beta_{1}=0.7$, $\beta_{2}=0.3$, $\beta_{3}=0.1$, $\beta_{4}=0.1 $, and $\beta_5=0.1$, because these weight settings do not adversely affect the convergence of the algorithm. These weights control the proportional contributions of different types of link information during intelligent routing optimization. For example, the weight with the highest value is $ \beta_{1} $, which means that when selecting a route, the main consideration will be to maximize the remaining bandwidth, while other link information will be used as secondary optimization conditions to affect the decision-making of the agent. This multi-objective optimization approach is adopted to overcome the shortcomings of single-objective optimization because it is better to comprehensively consider the current state of SDWN network traffic to improve the utilization of network resources. Meanwhile, the setting of the reward discount factors and the setting of the hyperparameters of the deep learning network will directly affect the convergence and stability of the intelligent algorithm; therefore, a related parameter analysis is given below.

\begin{figure*}[htp]
	\centering
	\begin{minipage}[t]{0.45\linewidth}
		\subfigure[discount factor $ \xi_{1},\xi_{2} $]{
			\centering
			\includegraphics[width=1\linewidth]{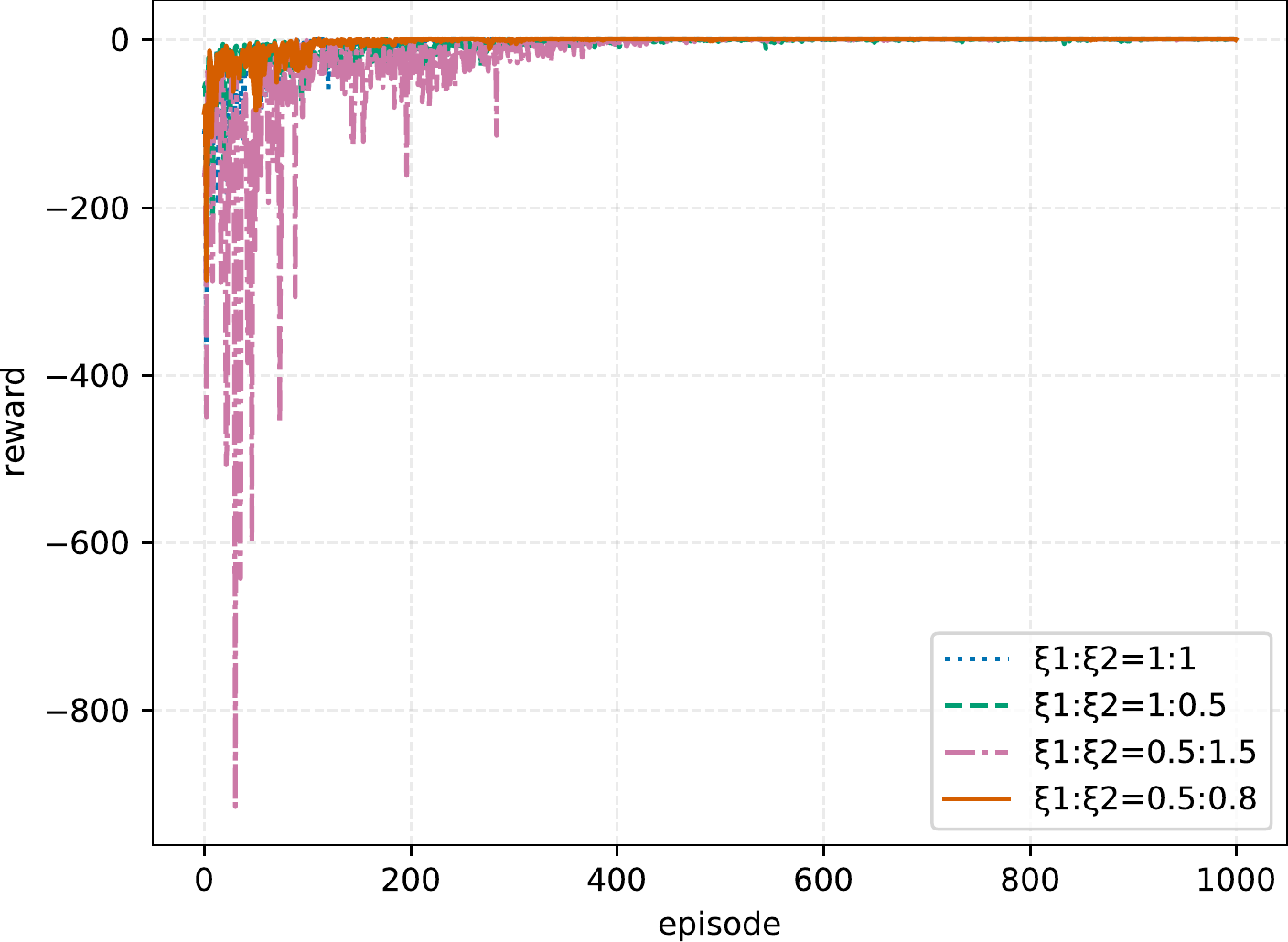}
			\label{subfig:11(a)}
		}
	\end{minipage}
	\hfill
	\begin{minipage}[t]{0.45\linewidth}
		\subfigure[ppo-update-time]{\centering
			\includegraphics[width=1\linewidth]{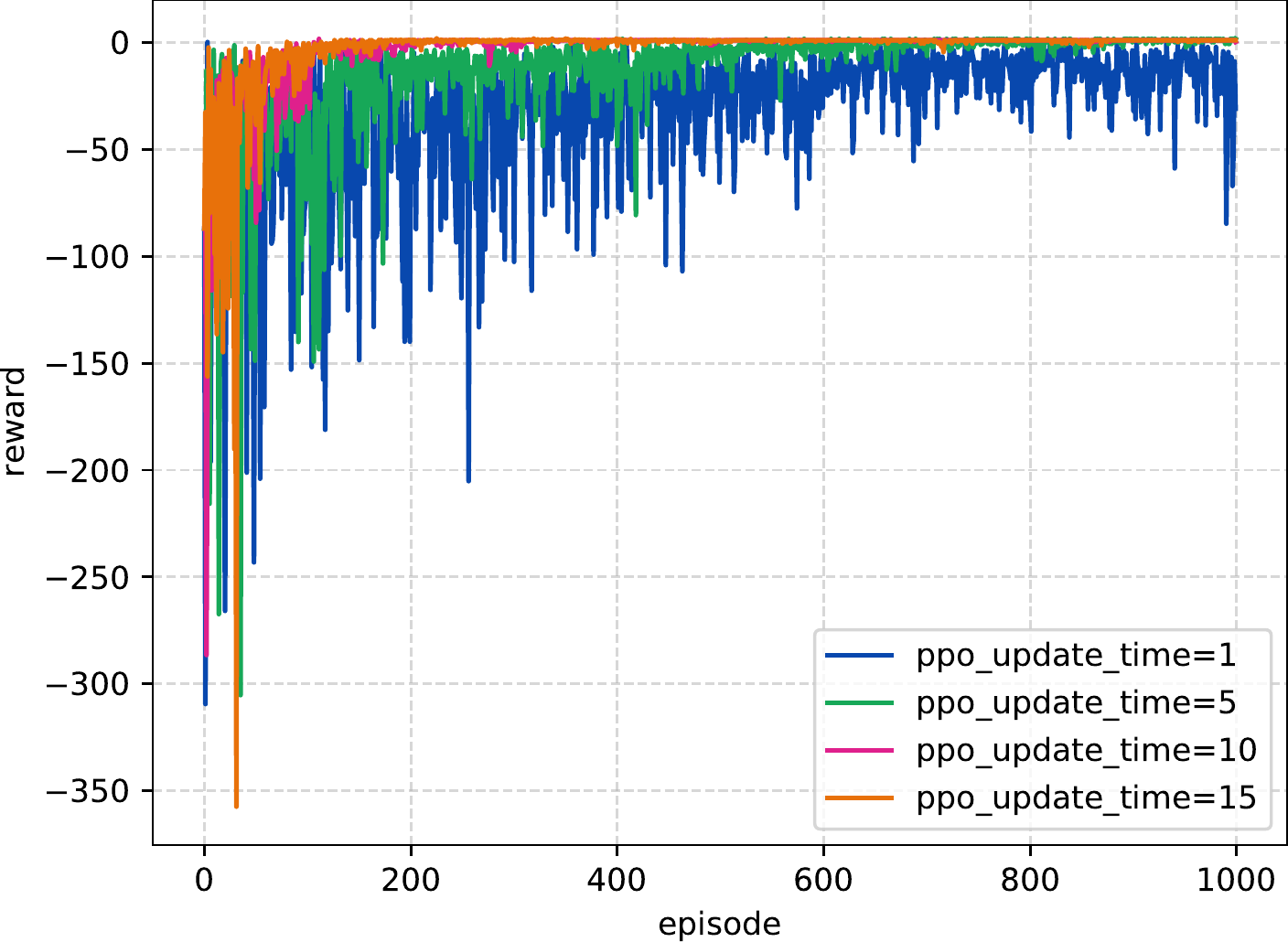}
			\label{subfig:11(b)}
		}
	\end{minipage}
    \begin{minipage}[t]{0.45\linewidth}
    	\subfigure[batch-size]{
    		\centering
    		\includegraphics[width=1\linewidth]{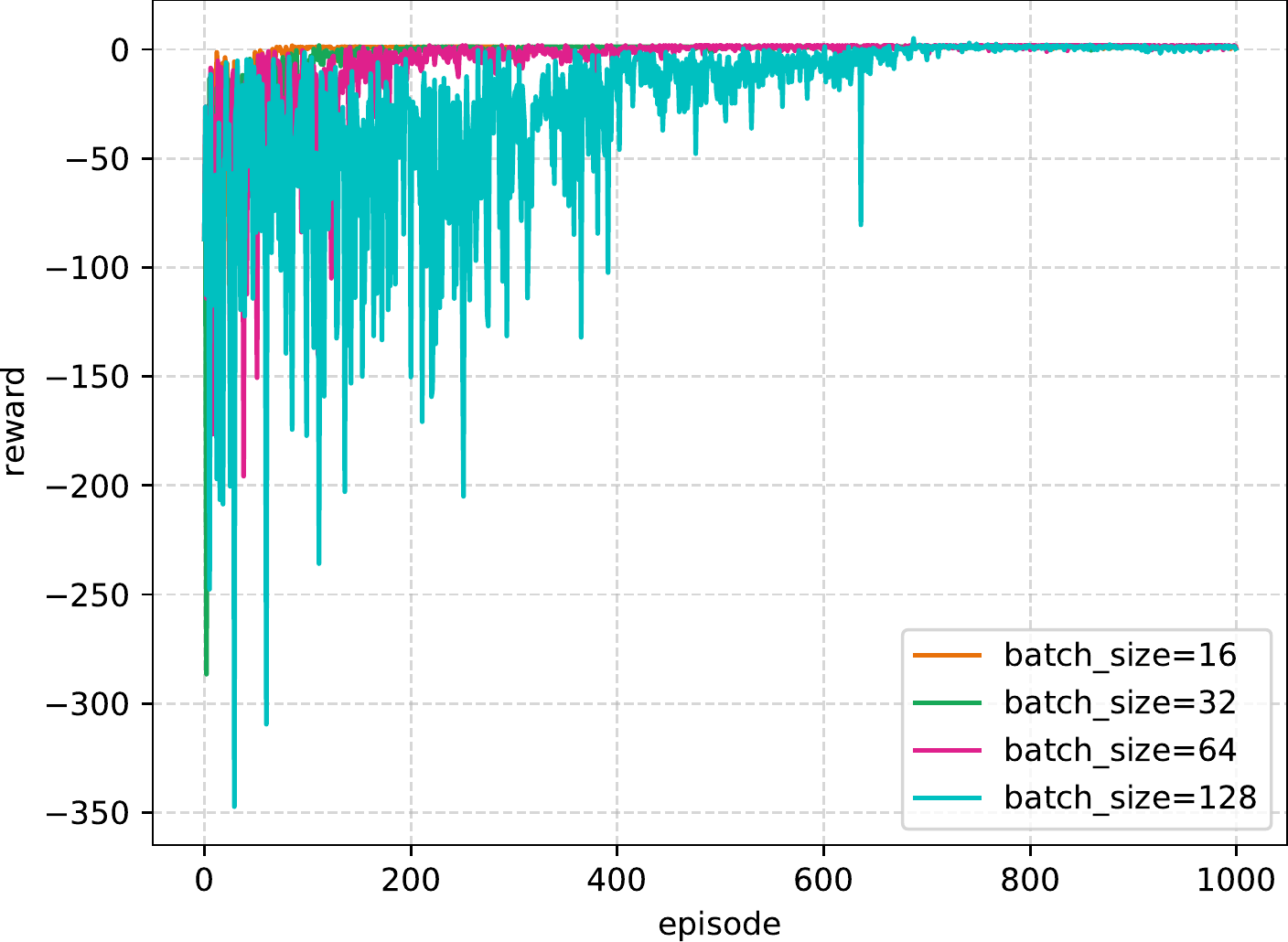}
    		\label{subfig:11(c)}
    	}
    \end{minipage}
    \hfill
    \begin{minipage}[t]{0.45\linewidth}
    	\subfigure[KL-Clip]{\centering
    		\includegraphics[width=1\linewidth]{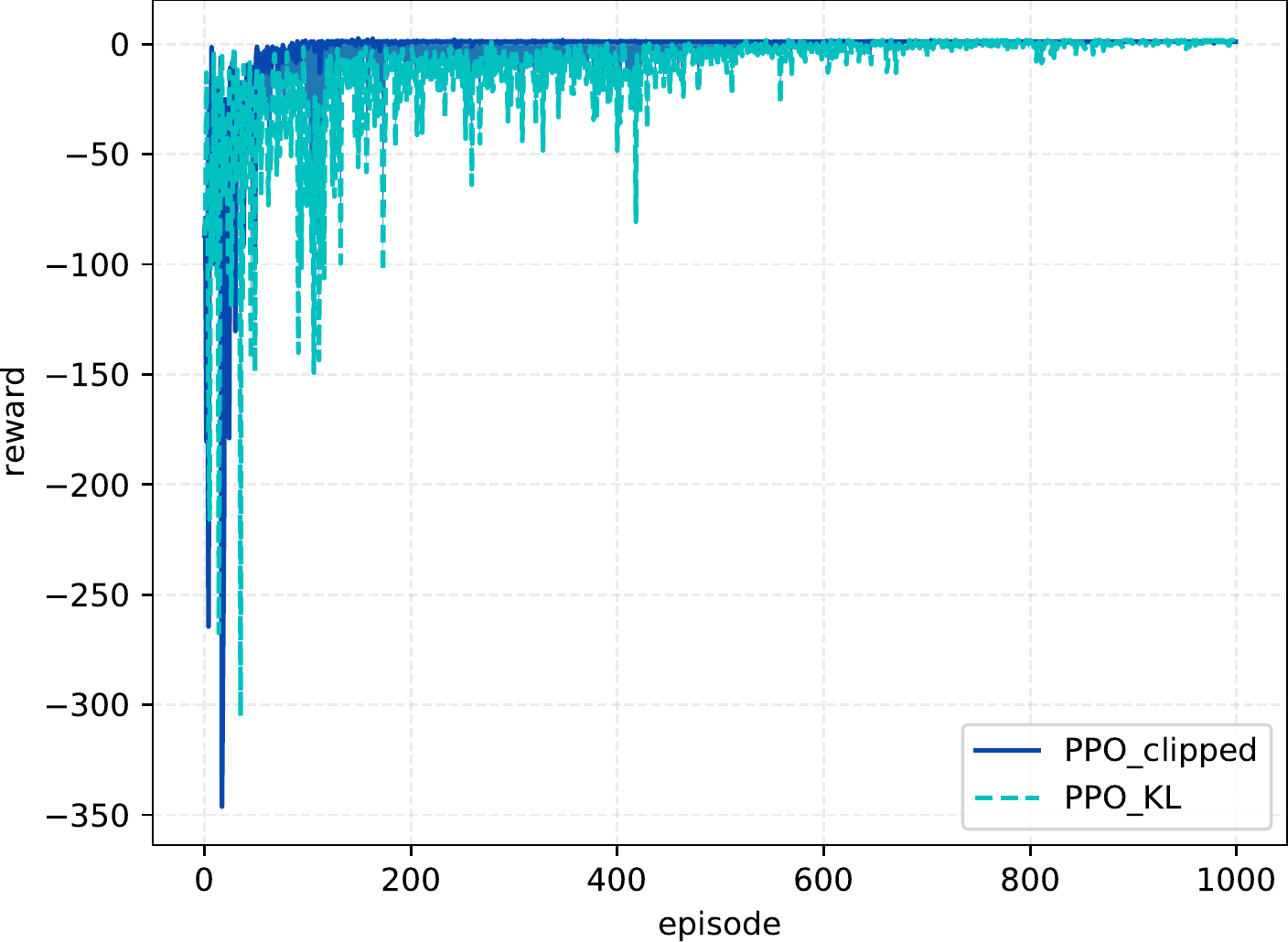}
    		\label{subfig:11(d)}
    	}
    \end{minipage}
	\caption{\textbf{Different parameter settings.}\label{fig11}}
\end{figure*}

The setting of the reward and penalty weights, $ \xi_{1} $ and $ \xi_{2} $, respectively, can have an impact on the direction of the agent's decisions and the convergence efficiency of the algorithm. Improper values for $ \xi_{1} $ and $ \xi_{2} $ can prevent the agent from learning the optimal path, which can affect the convergence of the algorithm and even lead to non-convergence. As shown in Fig. \ref{subfig:11(a)}, through multiple experiments, it has been found that setting the standard reward $R_s $ to $1$ and $  \xi_1:\xi_2=0.5:0.8 $ results in the fastest convergence speed with relatively stable waveforms, while  $ \xi_1:\xi_2=0.5:1.5 $ results in the slowest convergence speed, taking nearly 400 iterations to reach convergence.

The following analysis focuses on the impact of the algorithm hyperparameters on the reward received by the intelligent agent. The following hyperparameters are analyzed: the learning rate $ \alpha $, the batch size $k$, and the PPO network update frequency $f$. The $ \alpha $ determines when the objective function can converge to a local minimum and when it will converge to the global minimum. A suitable learning rate can make the objective function converge to a local minimum in a suitable amount of time. Since the PPO convolutional neural network consists of an actor network and a critic network, two different learning rates are set to adapt to the different network parameters and loss functions in the framework, namely, $ \alpha_{1} $ and $ \alpha_{2} $. As shown in Table \ref{tab4}, when $ \alpha_{1}=1e^{-3}$ and $\alpha_{2}=3e^{-3} $, the agent obtains the maximum reward, and the convergence effect is the best. At the same time, the agent takes the fewest steps when making routing decisions and seeks the globally optimal routing policy.

\begin{table}[htbp] 
	\renewcommand\arraystretch{1.5}
	\centering  
	\caption{\textbf{The impact of learning rate on algorithm convergence}}
	\label{table}
	\setlength{\tabcolsep}{6pt}
	\begin{tabularx}{\linewidth}{|>{\centering\arraybackslash}X|>{\centering\arraybackslash}X|>{\centering\arraybackslash}X|>{\centering\arraybackslash}X|}
		\hline
	   $  \alpha_{1},\alpha_{2} $	&  Reward&	Steps&	Episodes to reach convergence \\
		\hline 
      $1e^{-1},3e^{-1} $	&  $ -0.42 $ & $ 45 $ & $ 800 $\\
      $1e^{-2},3e^{-2} $	&  $ -0.82 $ & $ 34 $ & $ 290 $\\
       $ 1e^{-3},3e^{-3} $	&  $ 1.78 $ & $ 5 $ & $ 172 $\\
       $ 1e^{-4},3e^{-4} $	&  $ -180 $ & $ 478 $ & $ 890 $\\
       $ 1e^{-5},3e^{-5} $	&  $ -- $ & $ -- $ & $ -- $\\
       $ 1e^{-6},3e^{-6} $	&  $ -- $ & $ -- $ & $ -- $\\
		
		\hline
	\end{tabularx}
	\label{tab4}
\end{table}

The update frequency of the parameters in the PPO network affects the decision efficiency of the actor and critic networks. A suitable update frequency can enable the agent to make more accurate decisions and obtain higher reward values. In this experiment, we set the frequency $ f $ to 1, 5, 10, and 15. As shown in Fig. \ref{subfig:11(b)}, when the update frequency is set to 15, the agent obtains the maximum reward value, and the convergence effect is the best. Convergence is reached at approximately 170 iterations. However, when the update frequency is set to 1, the waveform exhibits large-amplitude jitter, which prevents the algorithm from converging.

The batch size $k$ determines the number of samples sent to the neural network at one time, i.e., the amount of data that the agent retrieves from the experience pool. An appropriate batch size can make full use of the GPU's parallel computing power, thereby improving the training efficiency of the algorithm and reducing the time required for model training. When the $k$ value is changed during training, the environment of the intelligent agent needs to be reinitialized to avoid the effects of the differences before and after this change. In this study, batch sizes of 16, 32, 64, and 128 are tested. As shown in Fig. \ref{subfig:11(c)}, when the batch size is $k=16$ or $k=32$, the agent obtains the highest reward values, and the convergence speed is the fastest, with a relatively stable convergence effect and small fluctuations. The agent can obtain the maximum reward value within 200 episodes. However, when the batch size is $k=128$, the agent takes approximately 700 episodes to converge, and the fluctuations are significant. Therefore, the batch size $ k $ is set to 32 in this paper to analyze the impact of the PPO-KL and PPO-Clip methods on the executed routing policy of the intelligent agent in importance sampling. 

As shown in Fig. \ref{subfig:11(d)}, when using the PPO-Clip optimization solution, the agent can converge faster and obtain the maximum reward value, while the convergence waveform is relatively stable. When using the PPO-KL optimization solution, the waveform shows obvious fluctuations due to the sensitivity of the KL divergence weight $\sigma$ to the data distribution and the difficulty of setting its value. For PPO-Clip, this restriction does not need to be considered. Therefore, using the PPO-Clip method can improve the performance of the algorithm.

\subsection{Performance Analysis}
\begin{figure}[t!]
	\centering
	\includegraphics[width=\linewidth]{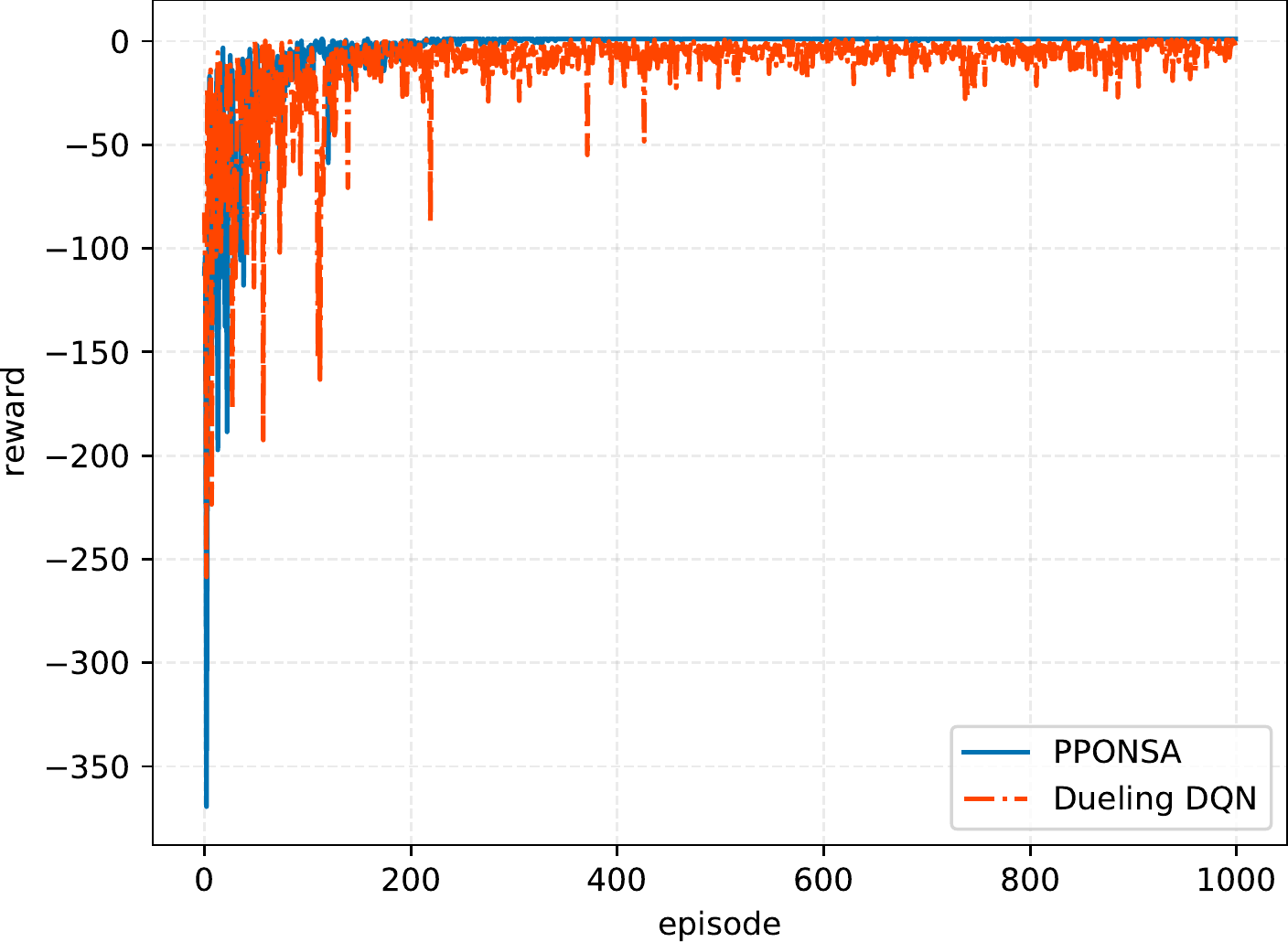}
	\caption{\textbf{ Comparison of convergence speed between PPONSA and Dueling DQN.}}
	\label{fig12}
\end{figure}

First, this section compares the performance of the DRL-PPONSA algorithm, a policy-based online DRL algorithm, with the value-based Dueling DQN algorithm. The policy-based approach shows better convergence and is more effective for problems with high-dimensional and continuous action spaces. It not only solves the challenges of model building for dynamic and diverse networks but also addresses the slow convergence and experience pool storage space issues of value-based reinforcement learning. To avoid any confounding influence of the prediction mechanism, this section analyzes experiments conducted using the predicted traffic matrix while keeping other parameter settings as consistent as possible. As shown in Fig. \ref{fig12}, the DRL-PPONSA algorithm reaches convergence at approximately 170 iterations, while the Dueling DQN algorithm takes approximately 300 iterations to converge; thus, DRL-PPONSA shows an overall improvement of 43.33\% in convergence speed. In the DRL-PPONSA algorithm, the prioritized experience replay scheme in Dueling DQN is replaced with the importance sampling technique. Importance sampling allows online training to be performed without storing the current sampled data; the entire experience pool is immediately deleted after the current data are used. In contrast, prioritized experience replay requires storing the data in the experience pool, and only some of those data are deleted when the capacity of the pool is exceeded, as this scheme relies on the extraction of historical experience data for offline training. Therefore, DRL-PPONSA can save computer memory, thus improving the convergence speed of the algorithm. 

Next, the DRL-PPONSA algorithm is compared with the classical Dueling DQN algorithm and the traditional OSPF, DVRP, and LSRP algorithms. The main performance indicators used to evaluate the algorithms include the average network throughput $ \overline{T } _{ij}  $, average network $ \overline{delay } _{ij} $ , average network packet loss rate $ \overline{loss } _{ij} $ , average network packet error rate $ \overline{pkt } _{err(ij)} $  , and average distance  $ \overline{distance } _{ij} $ of wireless access points. Equation \eqref{eq17} is used to calculate the network throughput, where $ e_{ij}\in E  $ represents one of the edges in the link and $ b_{e_{ij} }  $ represents the current amount of data sent on that link. Equations \eqref{eq18}-\eqref{eq22} gives the corresponding expressions for average network throughput, average network delay, and other metrics. 

\begin{equation}
{T_{ij}} = \sum\limits_{i,j = 0}^n {\frac{{b{w_{use\left( {{e_{ij}}} \right)}} \cdot \sqrt {\left( {1 - \left( {los{s_{{e_{ij}}}} + pk{t_{err\left( {{e_{ij}}} \right)}}} \right)} \right)}  \cdot b_{{e_{ij}}}}}{{2 \cdot dela{y_{{e_{ij}}}}}}} \label{eq17}
\end{equation}
\begin{equation}
{\overline T _{ij}} = \frac{1}{{{n_p}}}\sum\limits_{k = 1}^{{n_p}} {\sum\limits_{{e_{ij}} \in {p_{sd}}}^{} {{T_{ij}}} }.
 \label{eq18}
\end{equation}
\begin{equation}
{\overline {delay} _{ij}} = \frac{1}{{{n_p}}}\sum\limits_{k = 1}^{{n_p}} {\sum\limits_{{e_{ij}} \in {p_{sd}}}^{} {dela{y_{ij}}} }\label{eq19}
\end{equation}
\begin{equation}
{\overline {loss} _{ij}} = \frac{1}{{{n_p}}}\sum\limits_{k = 1}^{{n_p}} {\sum\limits_{{e_{ij}} \in {p_{sd}}}^{} {los{s_{ij}}} }\label{eq20}
\end{equation}
\begin{equation}
{\overline {pkt} _{err\left( {ij} \right)}} = \frac{1}{{{n_p}}}\sum\limits_{k = 1}^{{n_p}} {\sum\limits_{{e_{ij}} \in {p_{sd}}}^{} {pk{t_{err\left( {ij} \right)}}} }\label{eq21}
\end{equation}
\begin{equation}
{\overline {distance} _{ij}} = \frac{1}{{{n_p}}}\sum\limits_{k = 1}^{{n_p}} {\sum\limits_{{e_{ij}} \in {P_{sp}}}^{} {distanc{e_{ij}}} }\label{eq22}
\end{equation}
where $ p_{sd} $ represents the path from source node $ s\in \left \{ s_1,s_2,…,s_n  \right \} $ to destination node $ d\in \left \{ d_1,d_2,…,d_n \right \} $, $ n_p $ represents the number of pickles used to represent the amount of global network traffic obtained at time $t$.The traffic situation is collected for one day, and the average values are taken every three hours as the statistical results. The corresponding values are shown in the form of bar charts, and the experimental results are compared as follows.

\begin{figure*}[htp]
	\centering
	\begin{minipage}[t]{0.32\linewidth}
		\subfigure[Comparison of average network throughput.]{
			\centering
			\includegraphics[width=1\linewidth]{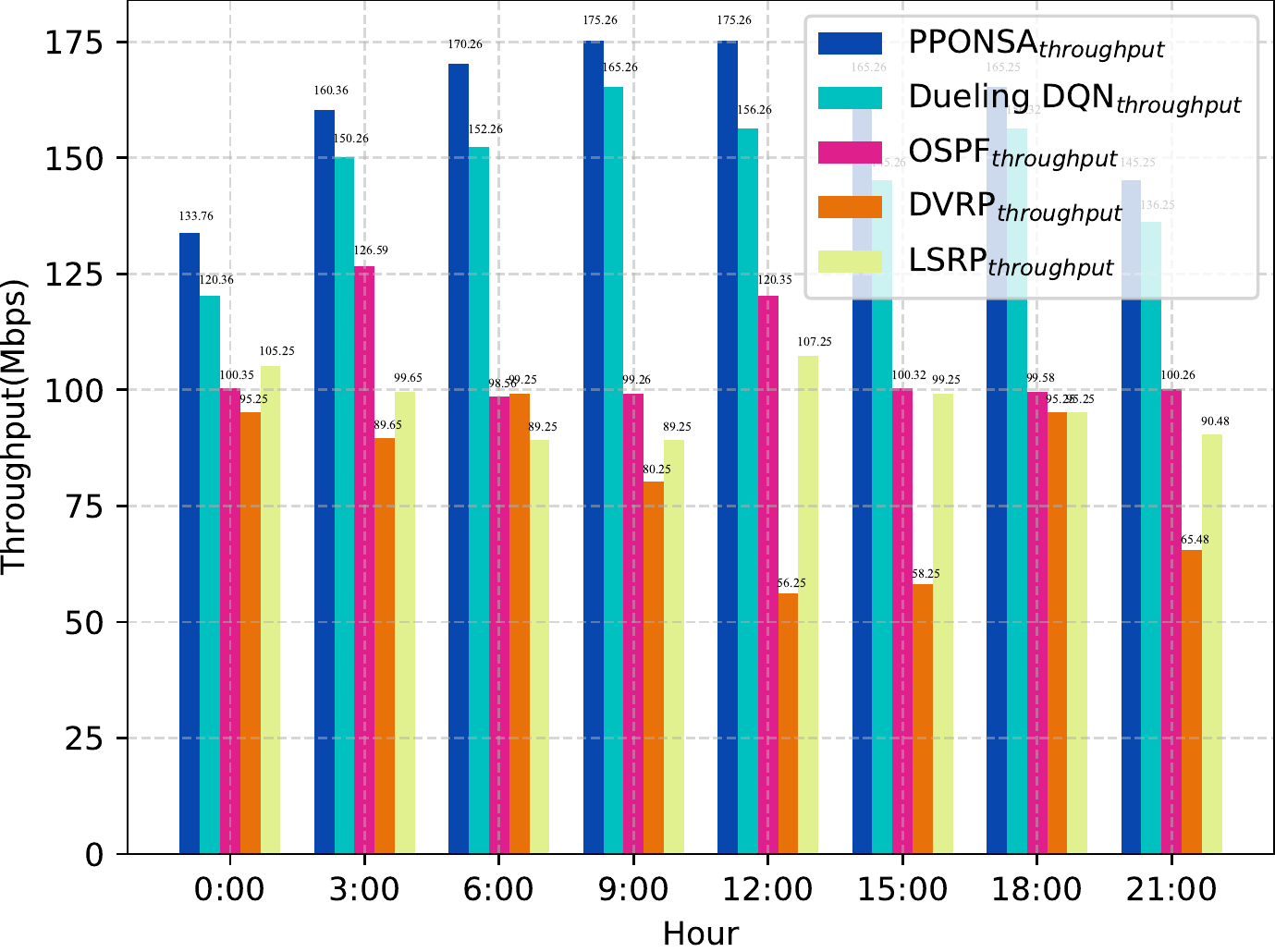}
			\label{subfig:13(a)}
		}
	\end{minipage}
	\begin{minipage}[t]{0.32\linewidth}
		\subfigure[Comparison of average network delay.]{\centering
			\includegraphics[width=1\linewidth]{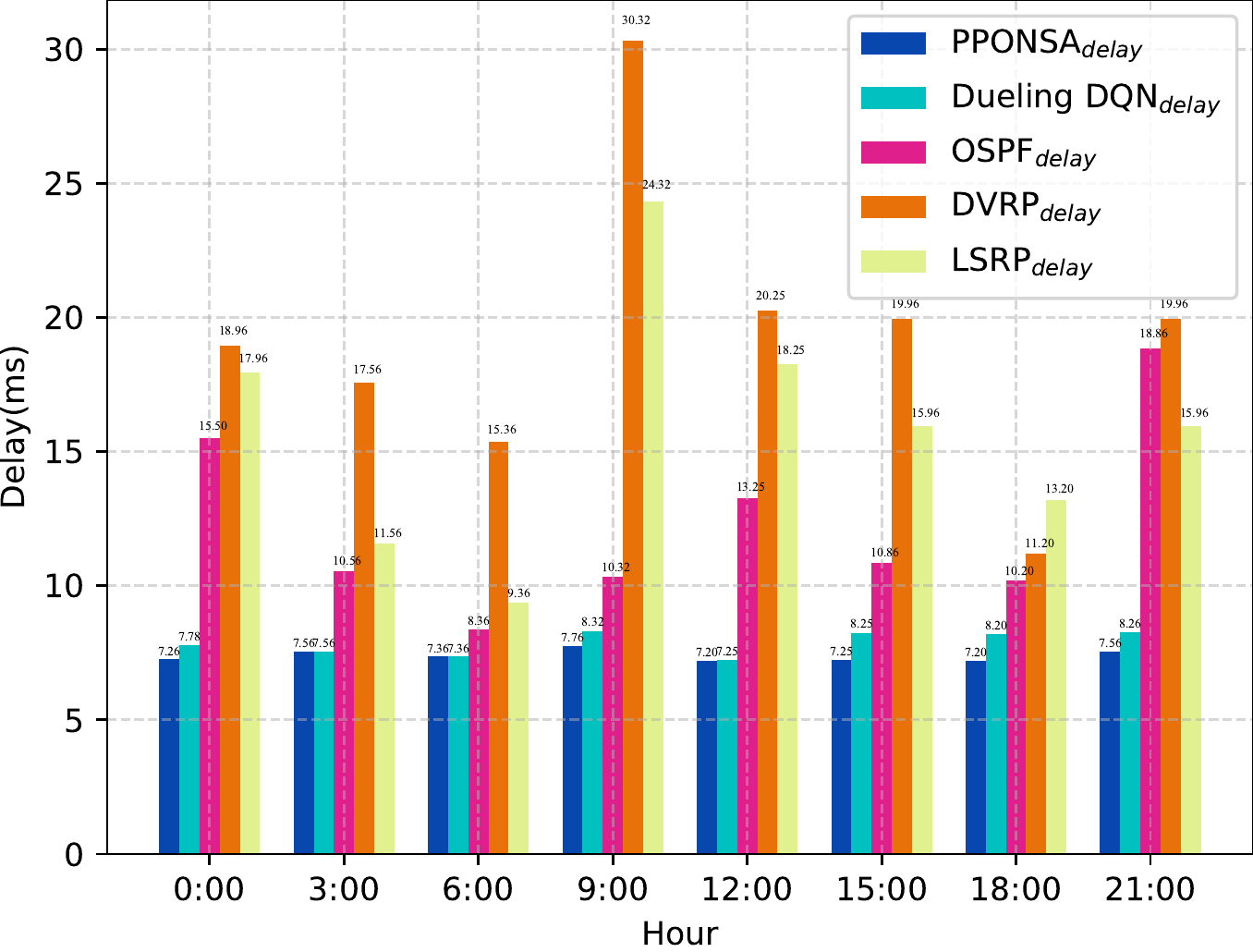}
			\label{subfig:13(b)}
		}
	\end{minipage}
	\begin{minipage}[t]{0.32\linewidth}
		\subfigure[Comparison of average packet loss rates.]{
			\centering
			\includegraphics[width=1\linewidth]{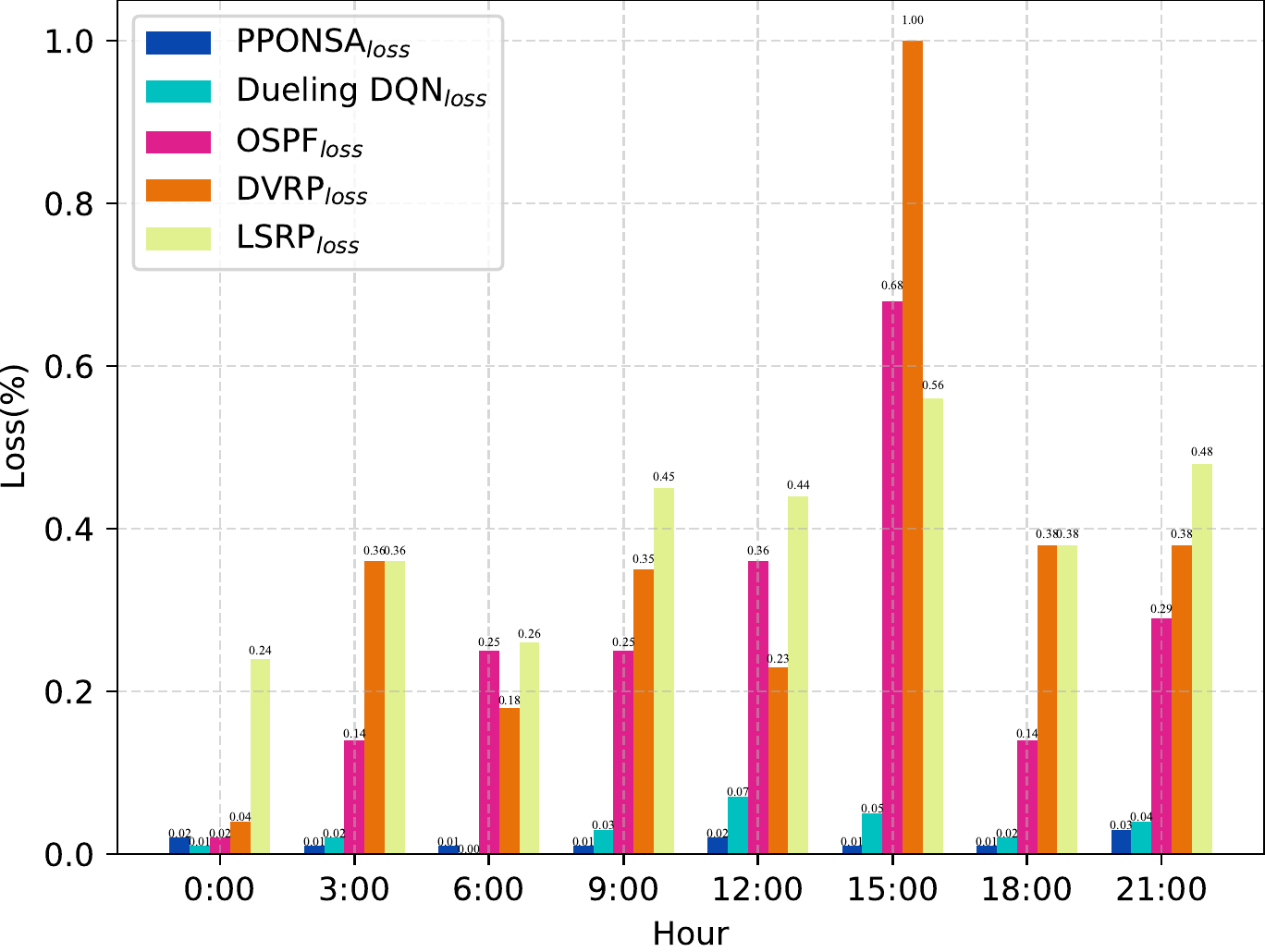}
			\label{subfig:13(c)}
		}
	\end{minipage}
	\hfill
	\begin{minipage}[t]{0.32\linewidth}
		\subfigure[Comparison of average packet error rates.]{\centering
			\includegraphics[width=1\linewidth]{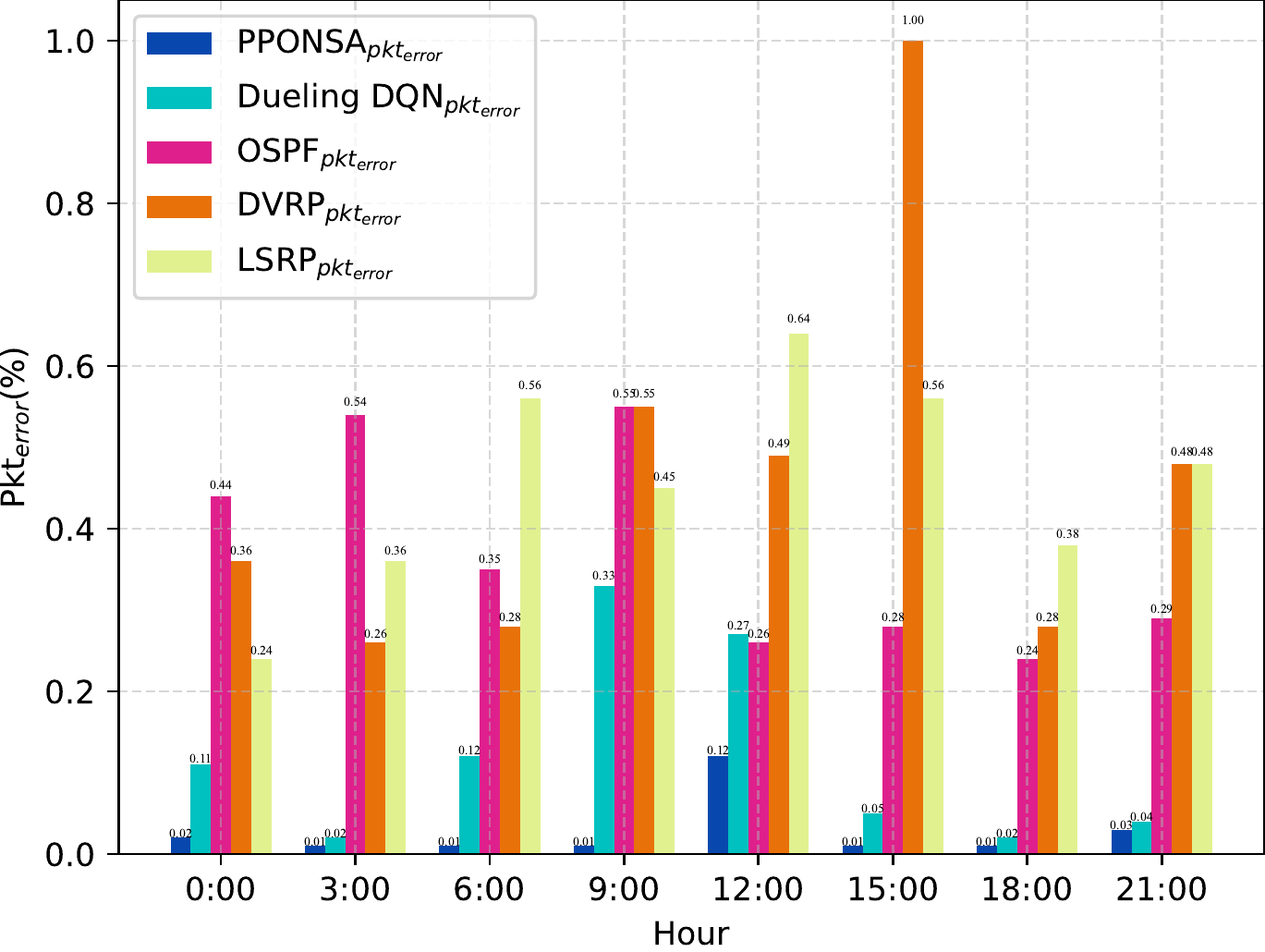}
			\label{subfig:13(d)}
		}
	\end{minipage}
    \begin{minipage}[t]{0.32\linewidth}
    	\subfigure[Comparison of average network link distances.]{\centering
    		\includegraphics[width=1\linewidth]{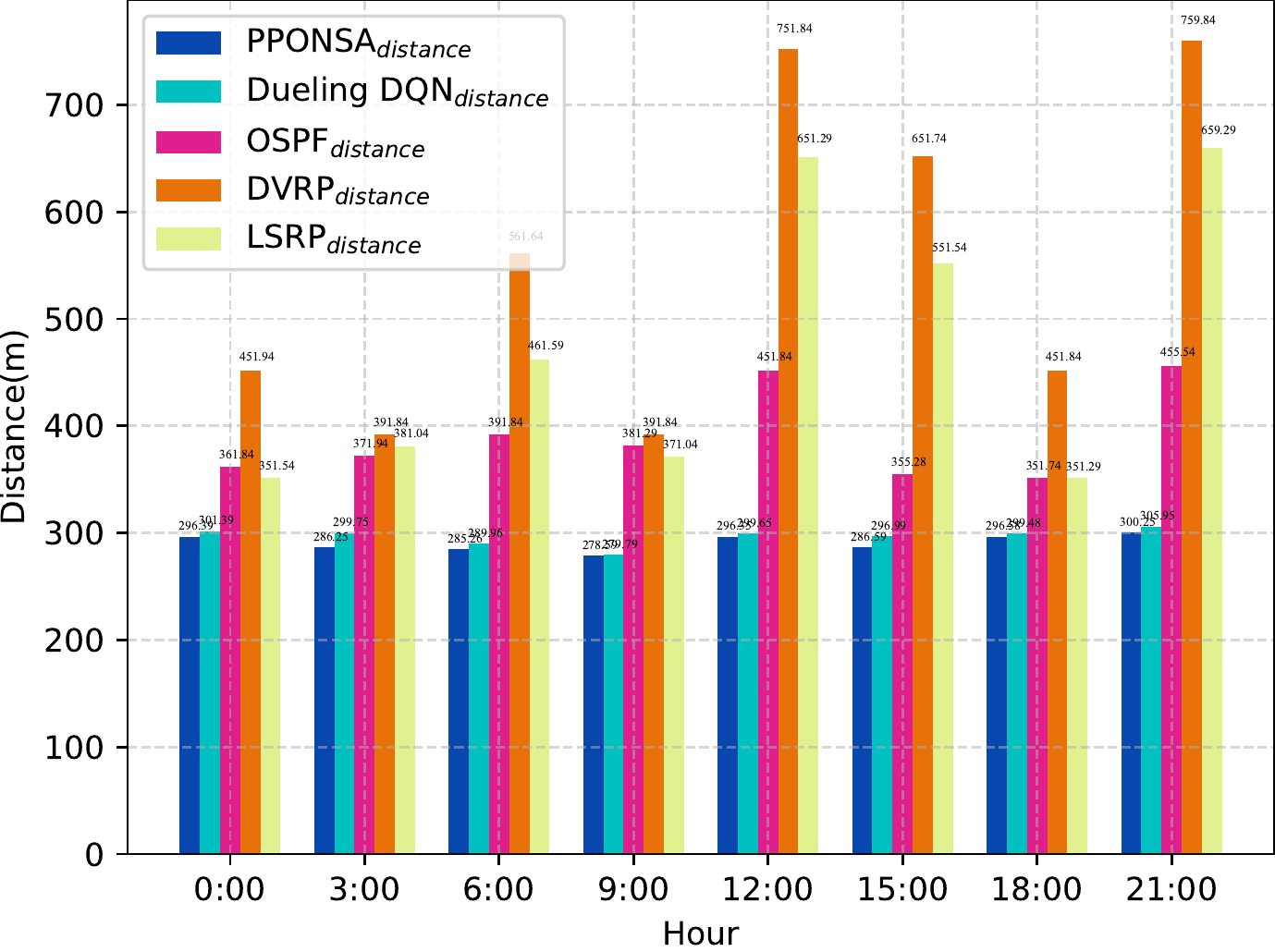}
    		\label{subfig:13(e)}
    	}
    \end{minipage}
	\caption{\textbf{Comparison of performance with other algorithms.}\label{fig13}}
\end{figure*}

The performance indicator shown in Fig. \ref{subfig:13(a)} measures the average total link throughput along the agent's path from the source node to the destination node. The DRL-PPONSA algorithm proposed in this paper achieves an average throughput that is 1.94\% higher than that of the Dueling DQN algorithm, 31.04\% higher than that of the OSPF algorithm, 47.81\% higher than that of the DVRP algorithm, and 36.72\% higher than that of the LSRP algorithm. The percentage improvements over the four compared algorithms, when the proposed algorithm performs best optimally, are 17.83\%, 47.66\%, 56.68\%, and 48.76\%, respectively. These findings indicate that the paths selected by the DRL-PPONSA algorithm have higher throughput, enabling the transmission of more data and better meeting the performance requirements for data transmission.

The measurement indicator in Fig. \ref{subfig:13(b)}  is the average delay on the agent's path from the source node to the destination node. The delay achieved with the DRL-PPONSA algorithm is superior to $Dueling\ DQN_{delay}$, $OSPF_{delay}$, $DVRP_{delay}$, $LSRP_{delay}$. Specifically, the average delay under DRL-PPONSA is 6.08\%, 39.58\%, 61.48\%, and 53.26\% lower, respectively, than the average delays under the other four algorithms. The average link delay results indicate that compared to the other algorithms, our algorithm is more inclined to search for paths with lower network latency, thereby avoiding network congestion and meeting the performance requirements of low-latency networks.

The measurement indicator in Fig. \ref{subfig:13(c)} is the average packet loss rate on the agent's path from the source node to the destination node. Compared to the average packet loss rates  $ Dueling\ {DQN_{loss}}$, $ OSPF_{loss}$,  $ DVRP_{loss}$, $ LSRP_{loss}$, and $ LSRP_{loss}$, the average packet loss rate under DRL-PPONSA is reduced by 50.84\%, 72.09\%, 89.79\%, and 82.07\%, respectively; thus, it is significantly lower than the packet loss rates of the other algorithms. In this experiment, the packet loss rate on each link is set between 0.1\% and 1\%, so each link will experience packet loss with a certain probability. When a link is congested, the probability of packet loss increases. From the experimental results, it can be seen that the link throughput on the path selected by the algorithm proposed in this paper is higher than that on the paths selected by the other algorithms, indicating that the proposed algorithm can meet the transmission needs of a high-traffic network while effectively avoiding packet loss.

The measurement indicator in Fig. \ref{subfig:13(d)} is the average packet error rate on the agent's path from the source node to the destination node. The DRL-PPONSA algorithm has an average packet error rate that is 77.08\%, 92.54\%, 89.79\%, and 94.05\% lower than $ Dueling\ DQN_{pkt_{err}}$, $ OSPF_{pkt_{err}}$, $ DVRP_{pkt_{err}}$, and $ LSRP_{pkt_{err}}$, respectively, significantly lower than those of the other algorithms. When a link experiences certain congestion, the probability of erroneous packets will increase. From the experimental results, it can be seen that the link delay and packet loss rate on the path selected by the algorithm in this paper are both smaller than those of the other algorithms, indicating that the proposed algorithm can meet the transmission needs of high-traffic networks while effectively avoiding packet errors.

The metric shown in Fig. \ref{subfig:13(e)} is the average link distance on the agent's path from the source node to the destination node. The DRL-PPONSA algorithm achieves average reductions of 19.68\%, 25.49\%, 47.75\%, and 38.43\% in link distance compared to $ Dueling\ DQN_{distance}$, $ OSPF_{distance}$, $ DVRP_{distance}$, $ LSRP_{distance}$, respectively. Although the average link distance of the proposed algorithm is similar to $ Dueling\ DQN_{distance} $, it can find markedly shorter transmission distances than the traditional algorithms can. In the optimization process, the distance between wireless APs is used as one of the measurement indicators to enable the intelligent agent to find shorter paths for data forwarding, thereby avoiding the selection of redundant paths and reducing the consumption of network bandwidth resources.

The experimental results show that the intelligent agent of the DRL-PPONSA algorithm proposed in this paper can search for the optimal routing policy based on the designed reward function. The convergence speed is significantly better than that of the Dueling DQN algorithm, and the average throughput is significantly higher than those of the traditional OSPF, DVRP, and LSRP algorithms. Meanwhile, the packet loss rate and transmission distance are significantly lower than those of the OSPF, DVRP, and LSRP algorithms. Although the proposed algorithm has shortcomings in terms of latency, it is still better than the traditional algorithms in the case of link congestion. The algorithm proposed in this paper is more stable in terms of transmission delay and more suitable for a network environment with small delay jitter. In environments with different volumes of network traffic, intelligent agents can adjust their reward functions based on multiple optimization objectives to dynamically update their routing policies to effectively ensure that the specified performance requirements are met during data transmission. The above experimental comparisons confirm that the SDWN-based intelligent routing algorithm proposed in this paper, DRL-PPONSA, not only shows good convergence but also offers good performance and stability.

\section{CONCLUSION}
This paper presents a novel method named DRL-PPONSA, for constructing intelligent routing paths in SDWN using DRL. The GCN-GRU prediction model is also incorporated in exploring unknown traffic information in the network, allowing the controller to perceive the network situation in real time and achieve intelligent control of the SDWN. In dynamic SDWN environments, DRL-PPONSA utilizes the policy gradient-based PPO reinforcement learning method to enable the agent to construct intelligent routing paths with higher bandwidth, shorter transmission distance, lower delay, and lower packet loss and error rates based on the measured network parameters. Moreover, the agent can also intelligently adjust the routing path and perform flow table installation when network parameters change. Comparative analysis with existing routing methods, the proposed intelligent algorithm exhibits good convergence and stability, significantly enhancing the overall performance and service quality of wireless networks.

Furthermore, the proposed algorithm is effective in addressing the dynamic network routing optimization problem. However, for large-scale and complex dynamic networks, relying on a single controller is insufficient to meet the performance requirements of the current network. Therefore, in future work, integrating a multi-controller-based multi-agent routing optimization method could achieve a more efficient solution to the SDWN routing.

\section*{Acknowledgment}
The author would like to thank the anonymous reviewers for their helpful comments.

\begin{IEEEbiography}[{\includegraphics[width=1in,height=1.25in,clip,keepaspectratio]{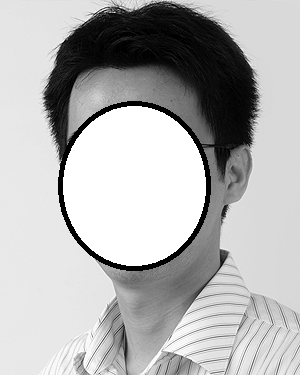}}]{JINQIANG LI} was born in 1997. He received his Ph.D. degree from Guilin University of Electronic Technology, Guilin, China, in 2021. His main research interests include reinforcement
	learning and software defined networking.
\end{IEEEbiography}

\begin{IEEEbiography}[{\includegraphics[width=1in,height=1.25in,clip,keepaspectratio]{author3.png}}]{Miao Ye} received his B.S degree in theory physics from Beijing Normal University in 2000 and his Ph.D. degree from School of Computer Science and Technology from Xidian University in 2006. He is currently a full professor and Ph.D. supervisor at Guilin University of Electronic Technology. His research interests include software defined networking, edge computing and edge storage, wireless sensor networks, deep learning.
\end{IEEEbiography}

\begin{IEEEbiography}[{\includegraphics[width=1in,height=1.25in,clip,keepaspectratio]{author3.png}}]{Linqiang Huang}  was born in
	1998. He is currently pursuing
	the master’s degree with the
	School of Computer Science
	and Information Security, Gui-
	lin University of Electronic
	Technology. His main research
	interests include reinforcement
	learning and software defined
	networking.
\end{IEEEbiography}

\begin{IEEEbiography}[{\includegraphics[width=1in,height=1.25in,clip,keepaspectratio]{author3.png}}]{Xiaofang Deng}   received the B.Eng. and M.Eng. degrees in communication engineering from the Guilin University of Electronic Technology (GUET), China, in 1998 and 2005, and the Ph.D. degree from the South China University of Technology (SCUT), Guangzhou, China, in 2016. She was a Visiting Scholar with Coventry University, in 2017. She is currently an Associate Professor with the School of Information and Communication Engineering, GUET. Her research interests include cognitive networks, network economy, and information sharing.
\end{IEEEbiography}

\begin{IEEEbiography}[{\includegraphics[width=1in,height=1.25in,clip,keepaspectratio]{author3.png}}]{Hongbing Qiu }  received his Ph.D. in information and communication engineering from Xidian University in 2004. From 1984 until now, he worked as a teacher in information and communication engineering in Guilin University of Electronic Technology. From 2012. 2 until 2012. 8, he has been a Visiting Researcher in University of Minnesota, Twin Cities, America. Professor Qiu is a doctoral supervisor at Xidian University and Guilin University of Electronic Technology, the director of the Ministry of Education Key Laboratory of Cognitive Radio and Information Processing, a member of Communication Theory and Signal Processing Committee of Communications Institute of China, a member of Broadband Wireless IP Standards Working Group, a member of Home Network Standards Working Group and more. His major research interests include mobile communication, wireless sensor networks and image signal processing.
\end{IEEEbiography}

\begin{IEEEbiography}[{\includegraphics[width=1in,height=1.25in,clip,keepaspectratio]{author3.png}}]{Yong Wang }  was born in 1964. He received his Ph.D. degree from East China University of Science and Technology, Shanghai, China, in 2005. He is currently a full professor and Ph.D. supervisor at Guilin University of Electronic Technology. His main research interests are cloud computing, distributed storage systems, software defined networks and information security.
\end{IEEEbiography}

\EOD

\end{document}